\documentclass[12pt]{article}
\usepackage{amsmath}
\usepackage{graphicx,psfrag,epsf}
\usepackage{enumerate}
\RequirePackage[colorlinks,citecolor=blue,urlcolor=blue]{hyperref}
\usepackage[shortlabels,inline]{enumitem}
\usepackage{natbib}
\usepackage{url} 
\usepackage{commath}

\usepackage{amsfonts}
\usepackage{amssymb}
\usepackage{xspace}
\usepackage{tabularx}
\usepackage{booktabs,tabulary}
\usepackage{multirow}
\usepackage{algorithm}
\usepackage{algpseudocode}
\usepackage{pifont}
\usepackage{subcaption}
\usepackage{float}
\setlist[enumerate, 1]{1\textsuperscript{o}}
\usepackage{mathtools}
\usepackage[font=small]{caption}
\usepackage{color}
\usepackage{xcolor,colortbl}
\usepackage{array}
\usepackage{bm}
\usepackage{lscape}
\usepackage{arash_macros}
\usepackage{bbm}
\usepackage[normalem]{ulem}
\useunder{\uline}{\ul}{}
\usepackage{diagbox}
\usepackage{setspace}

\newcommand{\blind}{0}

\newcommand{\thh}{\widehat\theta}
\newcommand{\rhoh}{\widehat\rho}
\newcommand{\Sigt}{\widetilde{\Sigma}}
\newcommand{\rhot}{\widetilde{\rho}}

\newcommand{\wc}{\check w}
\newcommand{\Eb}{\overline{E}}
\newcommand{\Sigh}{\widehat{\Sigma}}

\newcommand{\Expect}{{\rm I\kern-.3em E}}
\def\spacingset#1{\renewcommand{\baselinestretch}%
	{#1}\small\normalsize} \spacingset{1}

\newcommand{\beh}{\widehat \beta }
\newcommand{\sigh}{\widehat \sigma }
\newcommand{\pih}{\widehat\phi}
\newcommand{\Ch}{\widehat C}
\newcommand{\zh}{\widehat z}

\newcommand{\cond}{\,|\,}
\newcommand{\conf}{F}

\newcommand{\xnew}[1]{x_{#1,\text{new}}}
\newcommand{\ynew}[1]{y_{#1,\text{new}}}
\newcommand{\xtr}{x^{\text{train}}}
\newcommand{\ytr}{y^{\text{train}}}

\newcommand{\mmcl}[0]{\texttt{MMCL}\xspace}
\newcommand{\mmclpp}[0]{\texttt{MMCL++}\xspace}
\newcommand{\gmr}[0]{GMR\xspace}
\begin{document}



\if0\blind
{
\title{\bf Grouped Mixture of Regressions}
\author{Haidar Almohri$^\dagger$, PhD \\
	Ratna Babu Chinnam$^\dagger$, Ph.D., Professor\\
	Arash A. Amini$^\ddagger$, Ph.D, Professor
	\thanks{The authors gratefully acknowledge the support of \textit{Urban Science} for sponsoring this research.}\hspace{.2cm}\\
	$^\dagger$Department of Industrial and Systems Engineering\\
	Wayne State University, Detroit, MI 48201, U.S.A.\\[1ex]
	$^\ddagger$Department of Statistics\\
	 University of California, Los Angeles\\[1ex]
 {\normalsize \textit{\{Haidar.Almohri, Ratna.Chinnam\}@wayne.edu}}\\ {\normalsize \textit{AAAmini@ucla.edu}}}
\maketitle
} \fi

\if1\blind
{
\title{\bf Grouped Mixture of Regressions}
\maketitle
} \fi

\bigskip
\begin{abstract}
Finite Mixture of Regressions (FMR) models are among the most widely used approaches for dealing with heterogeneity in regression problems. One of the limitations of current FMR approaches is their inability to incorporate group structure in data when available. In some applications, it is desired to cluster groups of observations together rather than the individual ones. In this work, we extend the FMR framework to allow for group structure among observations, and call the resulting model the Grouped Mixture of Regressions (GMR). We offer a fast algorithm for estimating the model parameters using Expectation-Maximization (EM). We also show how the group structure can improve prediction by sharing information among members of each group, as reflected in the posterior predictive density under GMR.
The performance of the approach is assessed using both synthetic data as well as a real-world example.
\end{abstract}

\noindent%
\vfill

\newpage
\spacingset{1.45} 

\section{Introduction}
\label{sec:intro}
One of the challenges in modeling certain populations is that the observations might be drawn from different underlying processes. In such cases, a ``single" model may fail to efficiently represent the entire sample and as a result the accuracy and reliability of the model would suffer. This problem has been identified more than a hundred years ago~\citep{newcomb1886generalized, pearson1894contributions} and ``mixture models" were introduced in order to better account for the unobserved heterogeneity in the population. Since those early days, a lot of effort has gone into developing new methodologies and improving the existing models. In recent years, due to increasing availability and diversity of data, the topic has gained renewed interest among the researchers. Mixture models are being successfully employed in a variety of diverse applications such as speech recognition \citep{reynolds1995robust}, image retrieval \citep{permuter2003gaussian}, term structure modeling \citep{lemke2006term}, biometric verification \citep{stylianou2005gmm}, and market segmentation \citep{tuma2013finite}.

Among the family of mixture models, Finite Mixture of Regressions (FMR) models have been particularly popular in various fields and applications \citep{bierbrauer2004modeling, andrews2003retention, bar1978tracking}. This is mainly due to advantages of linear models such as simplicity, interpretability, and scientific acceptance. In FMR, it is assumed that the distribution of data can be represented using a convex combination of a finite $(K)$ number of linear regression models. Equivalently, each observation belongs to one of the $K$ classes, and given a class membership, it follows the regression model associated with that class. The difficulty is that the class memberships are not known in advance. 
%
%
%
%
\paragraph{Estimating the Parameters.}
FMR parameter estimation has been studied mainly from a likelihood point of view \citep{de1989mixtures}, with exceptions such as \cite{quandt1978estimating} where moment generating functions are used. The maximum likelihood approach using Expectation Maximization (EM) \citep{dempster1977maximum} remains the most widely used technique for estimating the parameters of the FMR. EM is an iterative procedure that is guaranteed to increase the likelihood at each step. As a by-product, one obtains approximate posterior distributions of the latent class memberships as well.
%
Other algorithms such as stochastic EM \citep{celeux1985sem} and classification EM \citep{celeux1992classification} have also been introduced as an attempt to improve the performance of the EM algorithm; see~\cite{faria2010fitting} for good discussion. 

\subsection{FMR with Group Structure}
Under the regular FMR setting, the response variable follows a mixture model where each component is a linear regression model based on the underlying covariates. In addition to the parameters of the regression models, FMR assigns a class membership to each observation along with a prior probability of belonging to each component. The result is equivalent to soft clustering of the observations into $K$ clusters, assuming $K$ components are employed. In some applications however, instead of individual observations, groups of observations are to be clustered or associated with the same component. For example, if the FMR is being employed to model data from a retail chain, it might be necessary to associate all observations stemming from any single store to the same component.

\begin{figure}[ht]
	
\begin{subfigure}{.5\textwidth}
\centering
\includegraphics[width=1\linewidth]{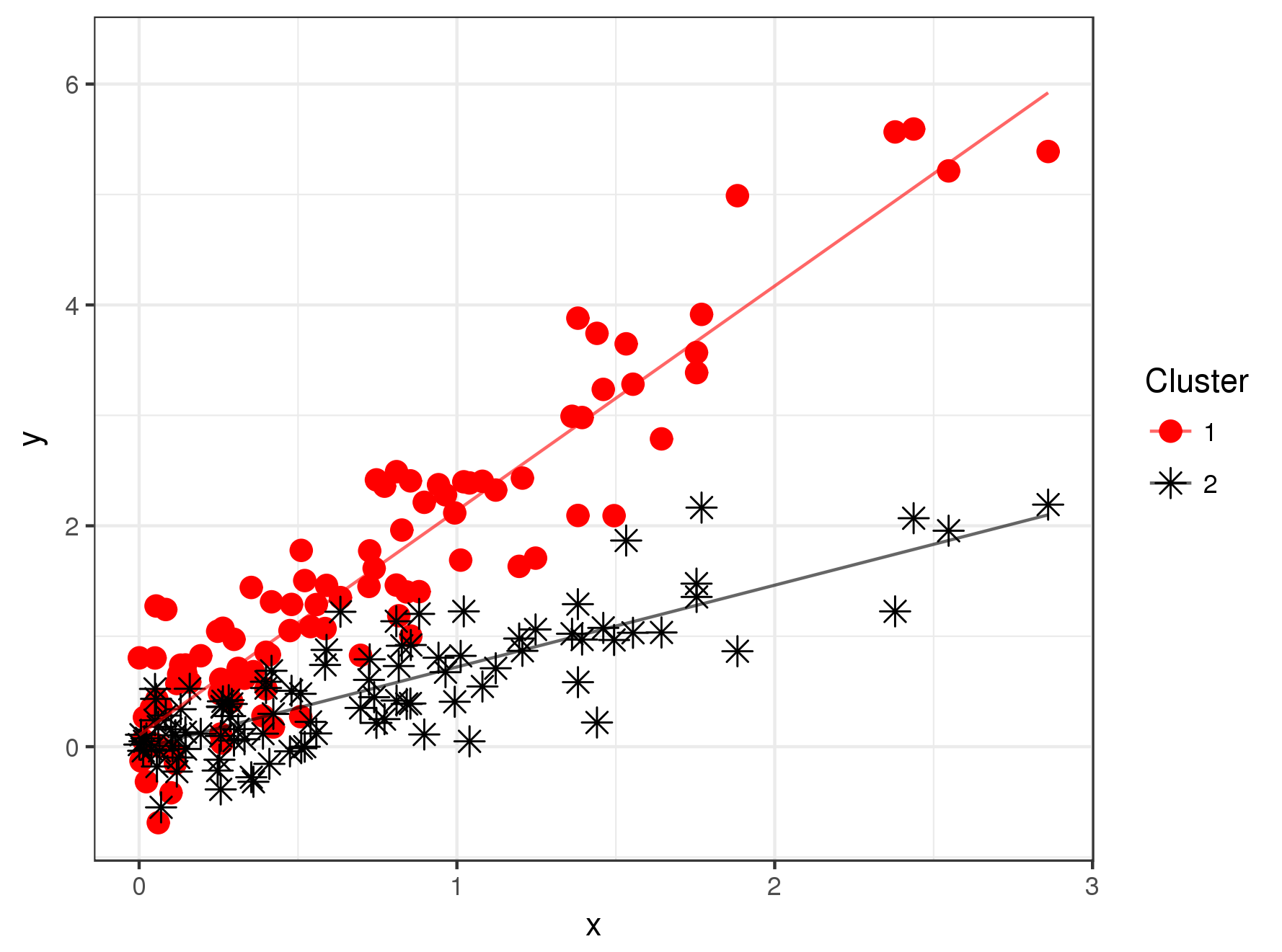}
\caption{\footnotesize{}}\label{fig:sfig1}
\end{subfigure}
\begin{subfigure}{.5\textwidth}
\centering
\includegraphics[width=1\linewidth, height = 0.75\linewidth]{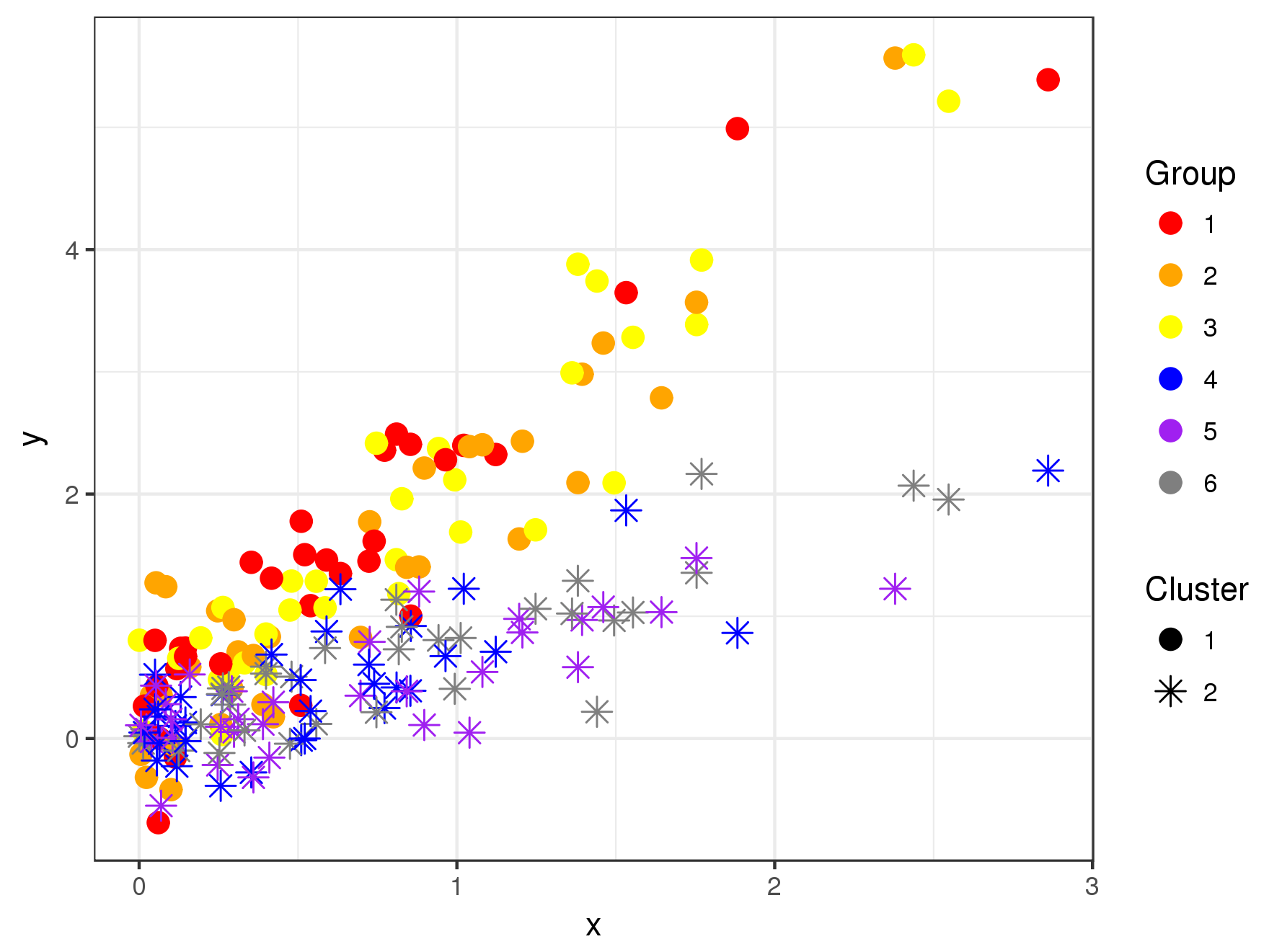}

\caption{\footnotesize{}}\label{fig:sfig2}
\end{subfigure}
\caption{FMR with ``group'' constraint: (a) Synthetic, two-component FMR without any constraint. (b) The same data divided into six groups where each group has to retain its data points.}
\end{figure}

This problem is similar to what is known as ``clustering with must-link constraint'', which was introduced by \cite{Wagstaff:2001}. 
The main idea is to utilize experts' domain knowledge prior to clustering process in order to obtain desired properties from the clustering solution. Figure \ref{fig:fig1} illustrates the concept. The data points are synthetically generated using two components: $y_1 = \frac{1}{2}x+\epsilon_1$ and $ y_2 = \frac{3}{4}x+\epsilon_2$, where $x \sim \mathcal{N}(0,1)$, $\epsilon_1 \sim \mathcal{N}(0,0.5)$, and $\epsilon_2 \sim \mathcal{N}(0,0.3)$. Figure~\ref{fig:sfig1} shows the linear relationship between the two groups ($y_1$ and $y_2$), without any grouping (must-link) structure. In Figure~\ref{fig:sfig2}, the data points are linked to create six groups (groups 1-3 belong to model $y_1$ and groups 4-6 to $y_2$). Data points with the same color belong to the same group. The desired outcome is to have all the data points in the same group end up having the same class membership. See \cite{basu2009constrained} for a good discussion of constrained clustering algorithms and applications.

\label{fig:fig1}

\cite{haidar:inpress} proposed a non-parametric, heuristic solution to the grouping problem called Mixture Models with Competitive Learning (\mmcl). They developed an iterative algorithm that given the current estimated component regression models, assigns each group to the cluster that best predicts the observations of that group in terms of the employed loss function. The component regression models are then updated by fitting a regression model to the aggregated observations assigned to each cluster. The procedure bears resemblance to the $k$-means clustering technique, with Euclidean distance replaced with prediction loss for the group. \cite{haidar:inpress} also provide an extension to \mmcl, called \mmclpp, that helps select the initial groups.


To the best of our knowledge, the work of~\cite{haidar:inpress} is the only attempt that addresses FMR with group structure in the literature. In contrast to their heuristic approach, we provide an exact parametric solution to this problem. We also derive an iterative likelihood-based parameter estimation algorithm for the model based on EM.

\subsection{Grouped Mixture of Regressions (\gmr)}\label{sec:model}
We assume that the observations belong to $R$ \emph{known groups}, denoted with labels $[R] := \{1,\dots,R\}$. In each group $r \in [R]$, we observe $n_r$ samples $(y_{ri},x_{ri}), i=1,\dots,n_r$ where $y_{ri} \in \reals$ is the response variable and $x_{ri} \in \reals^p$ is the vector of covariates or features. We will write $x_{rij}$ to denote the $j^{th}$ feature in the feature vector $x_{ri}$. For the most part, we will treat $x_{ri}$ as deterministic observations, i.e., we have \emph{fixed design} regression models.

We assume that there are $K$ latent (unobserved) clusters (data generating processes) and that all the observations in any group $r$ belong to a single cluster. Thus, we can assign a cluster membership variable $z_{r} \in \{0,1\}^K$ to each group $r \in [R]$. We will have $z_{rk} = 1$ iff group $r$ belongs to cluster $k$. With some abuse of notation, we also write $z_r = k$ in place of $z_{rk} = 1$.
%
%
%
%
Given the cluster membership variable $z_r$, we assume that group $r$ observations are independent draws from a Gaussian linear regression model with parameters specified by $z_r$, that is,
\begin{align}\label{eq:gauss:mixreg:model}
	p(y_{ri} \cond z_{r} = k) \stackrel{\text{indept}}{\sim } \mathcal{N}( \beta_k^T x_{ri}, \sigma_k^2), \; i =1,\dots,n_r,
\end{align}
where $\beta_k \in \reals^p$ is the coefficient vector of the $k$th regression model and $\sigma_k^2$ is the noise variance for component $k$. Note that we are assuming that the noise level only depends on the underlying cluster and not on the group. We write $\beta = (\beta_1 \mid \dots \mid \beta_K) \in \reals^{p \times K}$ and $\sigma^2 = (\sigma_1^2,\dots,\sigma_K^2) \in \reals^K$.


As is common in mixture modeling, we assume that $z_r$ follows a multinomial prior with parameter $\pi = (\pi_k)$, that is, $\pr(z_r = k) = \pi_k$ for $k \in [K]$, and $z_1,\dots,z_R$ are drawn independently.
%
The joint distribution of $y_r$ and $z_r$ is then given by:
\begin{align}
	p_{\theta}(y_r,z_r) = p_\theta(z_r) \prod_{i=1}^{n_r} p_\theta(y_{ri} \cond z_r) 
	= \prod_{k=1}^K \Big[\pi_k \prod_{i=1}^{n_r} p_\theta(y_{ri} \cond z_r = k)\Big]^{z_{rk}}
\end{align}
where we let $\theta = (\beta,\pi,\sigma^2)$ denote the collection of all the model parameters.
From~\eqref{eq:gauss:mixreg:model}, we have 
$p_\theta(y_{ri} \cond z_r = k) = \phi_{\sigma_k}\big( y_{ri} - \beta_k^T x_{ri} \big)$, where $\phi_\sigma(\cdot)$ is the density of the Gaussian distribution $\mathcal{N}(0,\sigma^2)$.
Therefore, the complete likelihood of $\theta$ given $(z,y)$ is:
\begin{align}\label{eq:likelihod}
	L(\theta \cond y, z)= p_\theta(y,z) = \prod_{r=1}^R p_{\theta}(y_r,z_r) = \prod_{r=1}^R \prod_{k=1}^K \Big[\underbrace{\pi_k \prod_{i=1}^{n_r} \phi_{\sigma_k}\big( {y_{ri} - \beta_k^T x_{ri}} \big)}_{ =: \; \gamma_{rk}(\theta)} \Big]^{z_{rk}}
\end{align}
The parameter $\gamma_{rk}(\theta)$ in~\eqref{eq:likelihod} is proportional (in $k$) to the posterior probability of $z_r$ given the observation $y_r$, that is, $p_\theta(z_r = k \cond y_r) \propto_k p_\theta(y_r,z_r = k ) = \gamma_{rk}(\theta)$. By normalizing $\gamma_{rk}(\theta)$ over $k$, we obtain the \emph{posterior probability of cluster assignments}:
\begin{align}\label{eq:posterior:memebership}
	p_\theta(z_r = k \cond y_r) = \frac{\gamma_{rk}(\theta)}{\sum_{k'} \gamma_{rk'}(\theta)} =: \tau_{rk}(\theta), \quad 
\end{align}
for any $k \in [K]$ and $r \in [R]$. 
We note that the overall posterior factorizes over groups, i.e., $p_\theta(z \cond y) = \prod_{r} p_\theta(z_r \cond y_r)$, so it is enough to specify it for each pair of $z_r$ and $y_r$. Thus, $\tau_{rk}(\theta)$ is the posterior probability that group $r$ belongs to cluster $k$, given all the observations $y$. These posterior probabilities are key estimation objectives. 

%
An estimate $\thh = (\beh, \pih, \sigh^2)$ of $\theta$ can be obtained by maximizing~\eqref{eq:likelihod}. The classical approach to performing such optimization is by the Expectation Maximization (EM) algorithm, the details of which will be given in Section \ref{sec:em}. Once we have an estimate $\thh$ of the parameters, we can calculate an estimate of the posterior probabilities as $\tau_{rk}(\thh)$.

\subsection{Posterior Prediction with \gmr}\label{sec:post:predict}
Now assume that we have a new test data point $(\ynew{r},\xnew{r})$ in group $r$, for which we observe only the feature vector $\xnew{r}$ and would like to predict $\ynew{r}$. Let $(\ytr,\xtr)$ denote all the observations used in the training phase. The common link between the training and test data points are the latent variables $z_1,\dots,z_R$. In other words, since we already have a good estimate of the membership of group $r$ based on the training data (via the posterior~\eqref{eq:posterior:memebership}), we can obtain a much better prediction of $\ynew{r}$ than what the prior model suggests. 
More precisely, the \emph{predictive density} for $\ynew{r}$ based on $\ytr$ is:
\begin{align*}
	p_\theta(\ynew{r} \cond \ytr) &= \sum_{z_r} p_\theta(\ynew{r} \cond z_r) \;p_\theta(z_r \cond \ytr).
\end{align*}
Since, $p_\theta(z_r = k \cond \ytr) = p_\theta(z_r = k \cond \ytr_r) =\tau_{rk}(\theta)$, we obtain the following estimate of the predictive density:
\begin{align}
	\begin{split}
		p_{\thh}(\ynew{r} \cond \ytr)
		&= \sum_{k=1}^K p_\theta(\ynew{r}\cond z_r=k)\, \tau_{rk}(\thh) \\
		&= \sum_{k=1}^K \tau_{rk}(\thh) \,\phi_{\sigh_k}\big( \ynew{r} - \beh_k^T \xnew{r} \big). \label{eq:predict:post}
	\end{split}
\end{align}
Note that $\thh$ is our estimate of the parameters based on the training data $(\ytr,\xtr)$.
In particular, the posterior mean based on~\eqref{eq:predict:post} is $\sum_{k=1}^K \tau_{rk}(\thh) \, \beh_k^T \xnew{r}$, which serves as the maximum a posteriori (MAP) prediction for $\ynew{r}$.

To summarize, since the membership group of the new observation is known, we obtain a predictive density of the form~\eqref{eq:predict:post} for new observations. Thus, we can utilize the group structure to leverage the information acquired during training phase when predicting new observations. This allows us to achieve a better prediction accuracy using the (posterior) latent cluster assignment. 
This type of information sharing between the training and test data does not occur in the usual FMR and is a unique strength of the proposed GMR model. In the usual FMR, the posterior mean predicted for a new data point will be the prior average of the mixture components: $\sum_{k=1}^K \pi_k \, \beh_k^T x_{\text{new}}$, and the only sharing that occurs between the training and test data is via the estimated parameters $\{\beh_k\}$.

\subsection{\gmr Parameter Estimation} \label{sec:em}
Let us now derive the EM updates for the model.
Recalling~\eqref{eq:likelihod}, the complete log-likelihood of the model is $	\ell(\theta \cond y, z) = \log p_\theta(y,z) = \sum_{r=1}^R \sum_{k=1}^K z_{rk} \log \gamma_{rk}(\theta)$, or 
\begin{align} \label{eq:loglik}
	\ell(\theta \cond y, z) = \log p_\theta(y,z) 
	&= \sum_{r=1}^R \sum_{k=1}^K z_{rk} \Big[\log \pi_k + \sum_{i=1}^{n_r} \log \phi_{\sigma_k}\big( y_{ri} - \beta_k^T x_{ri} \big) \Big].
\end{align}
Treating the class latent memberships $\{z_r\}$ as missing data, we perform the EM updates to simultaneously estimate $\{z_r\}$ and $\theta$:
\begin{description}[leftmargin=0.2cm, labelindent=.2cm,	itemsep=1pt, parsep=3pt, topsep=4pt]
\item[E-Step:] Replace~\eqref{eq:loglik} with its expectation under the approximate posterior of $\{z_r\}$: 
\begin{align} 
\begin{split} \label{eq:exp_lik}
	F(\theta;\thh) := E_{z \sim \tau(\thh)} [\ell(\theta \cond y, z)] &=  \sum_{r=1}^R \sum_{k=1}^K \tau_{rk}(\thh) \log \gamma_{rk}(\theta)\\
\end{split}
\end{align}
using $\ex_{z \sim \tau(\thh)}[ z_{rk}] = \tau_{rk}(\thh)$, 
where $\tau_{rk}(\theta)$ is the posterior given in~\eqref{eq:posterior:memebership}. 
\item[M-Step:] Maximize $F(\theta;\thh)$ over $\theta$, giving the update rules for the parameters $\theta = (\beta,\pi,\sigma^2)$.
\end{description}

\begin{algorithm}[t!]
	\setstretch{1.2}
	\caption{Grouped mixture of regressions (GMR)}\label{alg:gmr}
	\label{Palgorithm}
	\begin{algorithmic}[1]
		\State \makebox[3.25in][l]{Compute feature covariances for each group: }
		$\Sigh_r \gets \frac1{n_r}\sum_{i=1}^{n_r} x_{ri} x_{ri}^T$
		
		\State \makebox[3.25in][l]{Compute feature-response cross-covariances: }
		$\rhoh_r \gets \frac1{n_r}\sum_{i=1}^{n_r} y_{ri} x_{ri}$
		
		\State For any class posterior $\tau = (\tau_{rk})$ define the following weights:
		\begin{align*}
		\tau_{+k}(\tau) := \sum_r \tau_{rk}, 
		\quad w_{rk}(\tau) := n_r \tau_{rk},
		\quad w_{+k}(\tau) := \sum_r w_{rk},
		\quad \wc_{rk}(\tau) := \frac{w_{rk}}{w_{+k}}.
		\end{align*}
		and the weighted covariances: $\Sigt_k(\tau) := \sum_{r=1}^R \wc_{rk} \Sigh_r$ and 
		$\rhot_k(\tau) := \sum_{r=1}^R \wc_{rk} \rhoh_r$.

		\State For any parameter $\theta = (\pi,\beta,\sigma^2)$ and class posterior $\tau = (\tau_{rk})$, define the errors:
		\begin{align*}
		E_{rk}(\beta) := \frac1{n_r}\sum_{i}^{n_r}  (y_{ri} - \beta_k^T x_{ri})^2, \quad 
		\Eb_k(\beta,\tau) := \sum_r \wc_{rk}(\tau) E_{rk}(\beta)
		\end{align*}
		
		\While{not converged}
		\State \makebox[2.25in][l]{Update class frequencies: }
		\makebox[2in][l]{$\pi_k \gets \tau_{+k}(\tau)/R,$} $k \in [K] $
		
		\State \makebox[2.25in][l]{Update regression coefficients: }
		 \makebox[2in][l]{$\beta_k \gets \Sigt_k^{-1}(\tau) \,\rhot_k(\tau),$} $k \in [K] $
		
		\State \makebox[2.25in][l]{Update noise variances: }
		\makebox[2in][l]{$\sigma^2_k \gets \Eb_k(\beta, \tau),$} $k \in [K]$
		
		\State \makebox[2.25in][l]{Update class memberships:} 
		\makebox[2.in][l]{$\tau_{rk} \gets \tau_{rk}(\theta),$ as given in~\eqref{eq:posterior:memebership},} $r \in [R], k \in [K]$		

		\EndWhile
	\end{algorithmic}
	
\end{algorithm}

To derive the update rules, we maximize $F(\theta;\thh)$ by a sequential block coordinate ascent approach, in each step maximizing over one of the three sets of parameters $\pi, \beta$ and $\sigma^2$, while fixing the others. The updates are summarized in Algorithm~\ref{alg:gmr}. The details can be found in Appendix~\ref{sec:EM:details}.

\section{Empirical Analysis}
A Monte Carlo simulation study was performed to assess the quality of the proposed GMR algorithm. The results of this study is presented in this section.

\paragraph{Experiment setup.} We generate the synthetic data from the GMR model~\eqref{eq:gauss:mixreg:model} with a random design where the feature vectors are drawn as $x_i \sim N(0,\Sigma)$ given $\Sigma$. The covariance matrix $\Sigma$ is itself drawn from a normalized Wishart distribution. Recall that $K$ is the number of clusters (or mixture components) and $R$ the number of groups. In most of the experiments, we use equal number of observations per group, that is, $n_r$ is the same for all $r=1,\dots,R$. Letting $n = \sum_{r=1}^R n_r$ be the total number of observations, we have $n_r = n/R$. Let $G_k$ be the number of groups in cluster $k$. In general, $\sum_{k=1}^K G_k = R$; here, we take all $G_k$ to be equal so that $G_k = G := R/K$. Thus, it is enough to specify $n, G$, and $K$. Table~\ref{tab:exp} summarizes various setups used in our simulations.
We recall that $p$ is the dimension of the feature vectors $x_i$ and ``the noise level'' is $\sigma_k$ in~\eqref{eq:gauss:mixreg:model}. In each case, the number of groups $R$ and the number of observations per group $n_r$ is determined by the number of clusters $K$, number of groups in each cluster $G$, and total number of observations $n$. 
 For example, for $n = 800$, $G=10$, and $K=2$, we have $R = 20$ and $n_r = 40$. 
\begin{table}[!t]
\small
\centering
\caption[Monte Carlo Simulation Parameters (GMR)]{Monte Carlo Simulation Parameters}\label{tab:exp}
\begin{tabular}{|c|c|c|c|c|c|}
\hline
$K$ & $p$   & $G$          & $n$                      & Noise Level $(\sigma_k)$           & $\beta$-distance ($\delta_\beta$)        \\ \hline
2 & 2   & \multirow{2}{*}{10} & \multirow{2}{*}{(100, 200, 400, 800)} & \multirow{2}{*}{(2, 4, 6, 8, 10)} & \multirow{2}{*}{(4, 8, 12)} \\ \cline{1-2}
4 & (2, 4) &           &                       &                  &               \\ \hline
\end{tabular}
\end{table}

	
To study the effect of heterogeneity among regression coefficient vectors $\beta_k, k \in [K]$, we take $\beta_k$s to be equidistant points on a hypersphere in $\reals^p$ and vary their common distance, which we term $\beta$-distance and denote as $\delta_\beta$. More precisely, we will have $\norm{\beta_k} = \norm{\beta_\ell}$ and $\norm{\beta_k - \beta_\ell} = \delta_\beta$ for all $k \neq \ell$. Generating $\beta$s this way enables us to effectively compare the estimation errors among different runs of the experiment. The comparison can be carried out across different setups by normalizing the calculated error by $\delta_\beta$. Three values of $\delta_\beta$ that are found to be adequate for our experiments are also listed in Table \ref{tab:exp}.
Obviously, the smaller the distance (the smaller the $\beta$s) the harder the cluster separation.
	
The above equidistant setup is designed so that the data points are not easily separable in the input or output spaces, i.e. solely based on the $X$ or $y$ values. The degree of separation is only controlled by $\beta$-distance ($\delta_\beta$) while the noise level ($\sigma_k$) controls the uncertainty in relation between $X$ and $y$. Figure~\ref{fig:data} shows samples of the generated data for different scenarios. Note from the Figure that the clusters are not identifiable in $X$ or $y$ domains, whereas plotting $y$ against $X$ reveals the two clusters.
\begin{figure} [ht]
	\centering
\includegraphics[width=.8\linewidth]{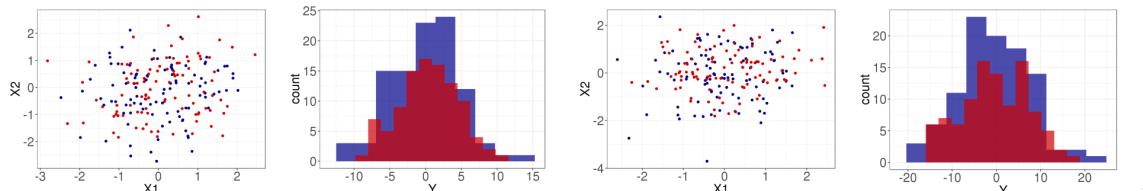}
\includegraphics[width=.8\linewidth]{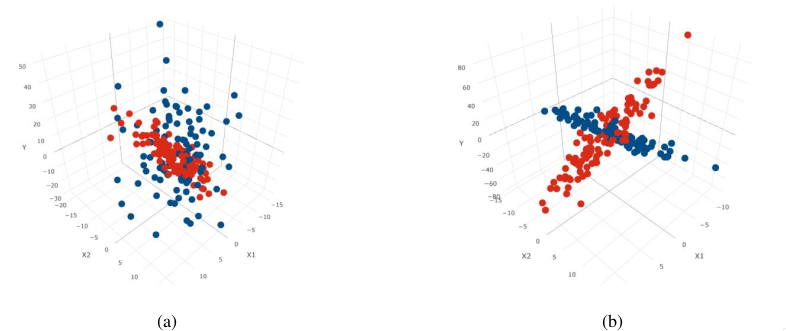}
\caption[Sample of the generated data for simulation]{Sample of the generated data for simulation for the case $p = k = 2$: Covariates $X$ (top left); the response values $y$ (top right); 3d plot for the $X$ and $y$ (bottom): (a) $\delta_\beta=4$, (b) $\delta_\beta=12$ } 
\end{figure}\label{fig:data}

\paragraph{Evaluation criteria.}
The Monte Carlo simulations are repeated 250 times for each pair of $\beta$-distance and the noise level as well as pairs of $p$ and $K$. This setup is maintained in all the experiments that will be discussed later in the manuscript. Four criterion are used to benchmark the performance of the algorithm: 
\begin{enumerate*}[label=(\arabic*)] 
	\item Normalized mutual information (NMI) for assessing the clustering accuracy,
	\item average $\beta$ estimation error,
	\item root mean squared error (RMSE) of prediction to assess the prediction power of the models, and
	\item the number of iterations to study the rate of convergence and the speed of the algorithms.
\end{enumerate*} 

NMI is a widely used measure for evaluating the quality of clustering algorithms when the true labels are available. Advantages of using NMI is its invariance to cluster label switching, and the aggressive penalization of the partitions close to random (relative to the true one). NMI is bounded between zero and one. The closer the value to zero, the higher the indication that the cluster assignments are largely independent, while a NMI close to one shows substantial agreement between the clusters. 

``$\beta$ estimation error'' is used as another measure of goodness of fit. We calculate this error by considering both the distance between the true and estimated $\beta$s, as well as the miss-classification error. More precisely, to each group $r$, we can assign two regression coefficient vectors, the estimated one $\beh^{(r)}$, and the true one $\beta^{(r)}$; $\beh^{(r)}$ is equal to $\beh_k$ if we have estimated group $r$ to be in cluster $k$. Similarly, $\beta^{(r)}$ is equal to $\beta_k$ if group $r$ is in true cluster $k$. We can define the average $\beta$ estimation error as:
\begin{align}
\label{eq:avg:beta:err}
\text{avg err}_{\beta} :=\frac1R \sum_{r=1}^R \vnorm{\beh^{(r)}- \beta^{(r)}}^2= \text{tr}(D^T \conf)
\end{align}
where $D=\big(\vnorm{\beh_k-\beh_\ell}^2, \; k, \ell \in [K] \big)$ is the $K\times K$ matrix of pairwise squared distances between $\beh_k$s, and $\conf$ is the confusion matrix between the estimated and true labels. The details for the second equality can be found in Appendix~\ref{sec:avg:beta:derivation}.

Prediction RMSE is obtained by designating a hold-out (or test) set and using the trained models to predict the responses over the hold-out set. 
In each simulation run, 80\% of the observations in each group is used for training the model and 20\% is held out to assess the prediction power. This setup is maintained in all the experiments that will be discussed later in the manuscript.

\section{GMR Results}
In this section we report in detail the results from the simulation and modeling experiments. Each factor of the study is presented in a subsection.

\begin{figure} [ht]
    \centering
    \begin{subfigure}[b]{0.49\textwidth}
      \centering
      \includegraphics[width=\textwidth]{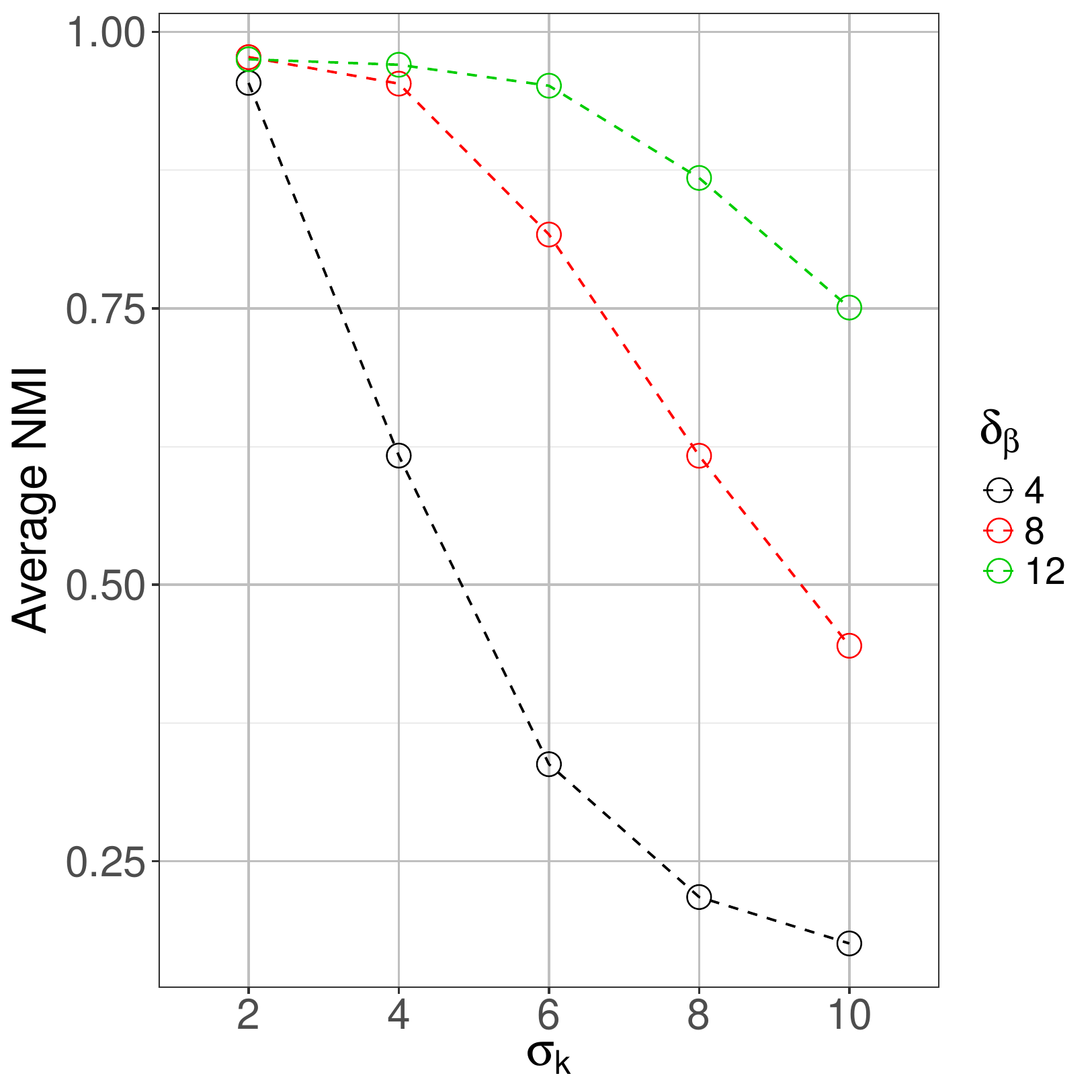}
      \caption{\footnotesize{}}\label{sfig:nmi_n2}
    \end{subfigure}
    \begin{subfigure}[b]{0.49\textwidth}
      \centering
      \includegraphics[width=\textwidth]{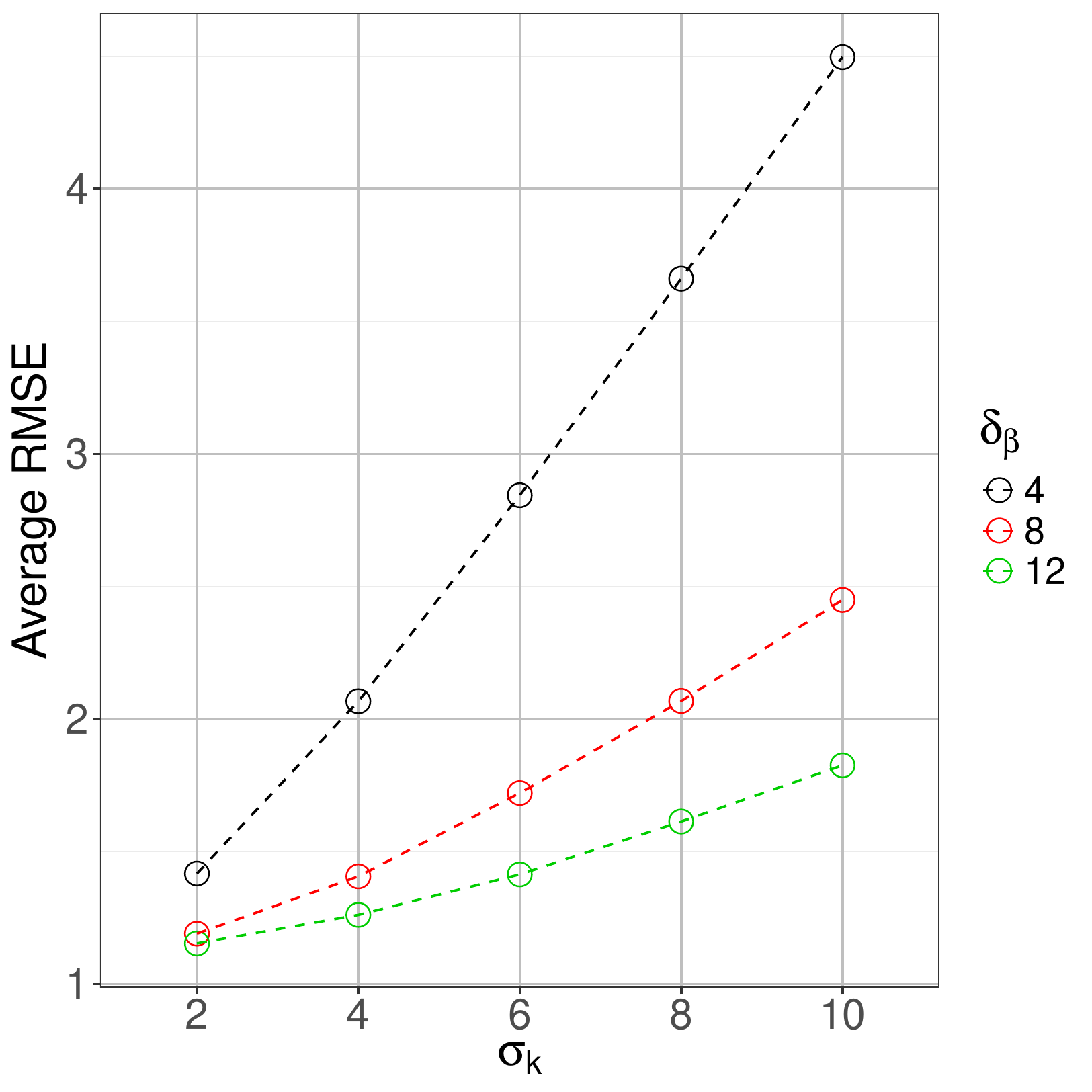}
      \caption{\footnotesize{}}\label{sfig:rmse_n2}
    \end{subfigure}
    \caption{The effect of $\delta_\beta$ and $\sigma_k$ for the case $n=200, K=4, p=4$; each colored line in a plot represents different value of $\delta_\beta$, $x$ axis shows different values of $\sigma_k$, and $y$ axis shows: (a) average NMI, (b) average RMSE for prediction.}
  \end{figure}

\paragraph{$\beta$-Distance ($\delta_\beta$) and Noise Level ($\sigma_k$).}
Figure~\ref{fig:bet:noise2} is the result of running the experiments for the setup $n = 200$, $p =4$, and $K =4$. Referring to Figure~\ref{fig:bet:noise2}, we observe that increasing $\sigma_k$ (decreasing the signal to noise ratio) leads to a drop in the performance of the algorithm. This is also the case with $\delta_\beta$, where we notice that the more separable the true $\beta$s are, the easier it is to estimate. We notice that at noise level $\sigma_k =10$, and $\delta_\beta = 4$, NMI (Figure~\ref{sfig:nmi_n2}) is close to zero, indicating that most of the times the algorithm fails to recover the true clusters. Similar trends are observed for the setup $n = 100$, $p =2$, and $K =2$, as illustrated by Figure~\ref{fig:bet:noise} in Appendix~\ref{sec:tables}.
%
  \label{fig:bet:noise2}

\begin{figure} 
    \centering
    \begin{subfigure}[b]{0.32\textwidth}
      \centering
      \includegraphics[width=\textwidth]{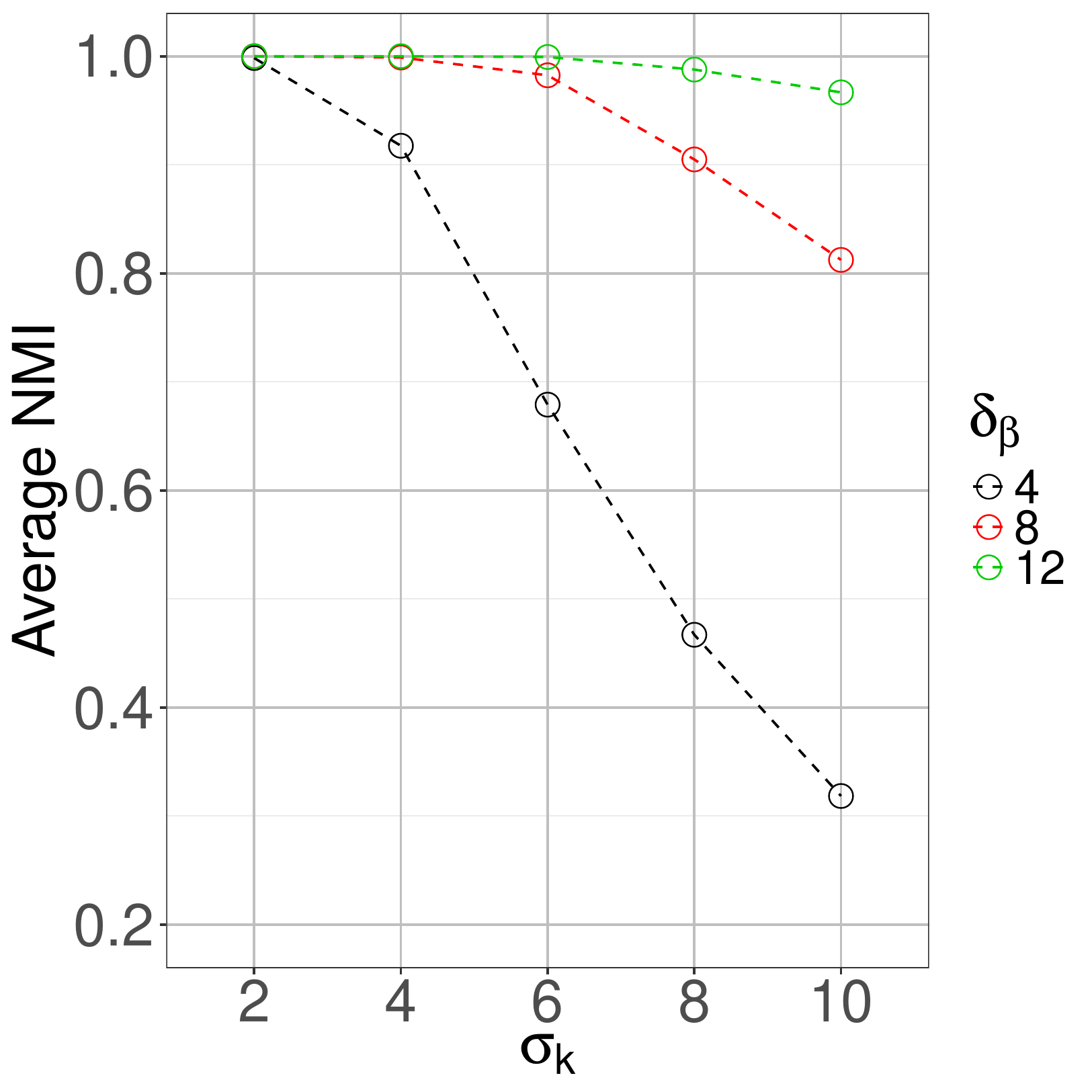}
      \caption{\footnotesize{}}\label{sfig:nmi_pk1}
    \end{subfigure}
    \begin{subfigure}[b]{0.32\textwidth}
      \centering
      \includegraphics[width=\textwidth]{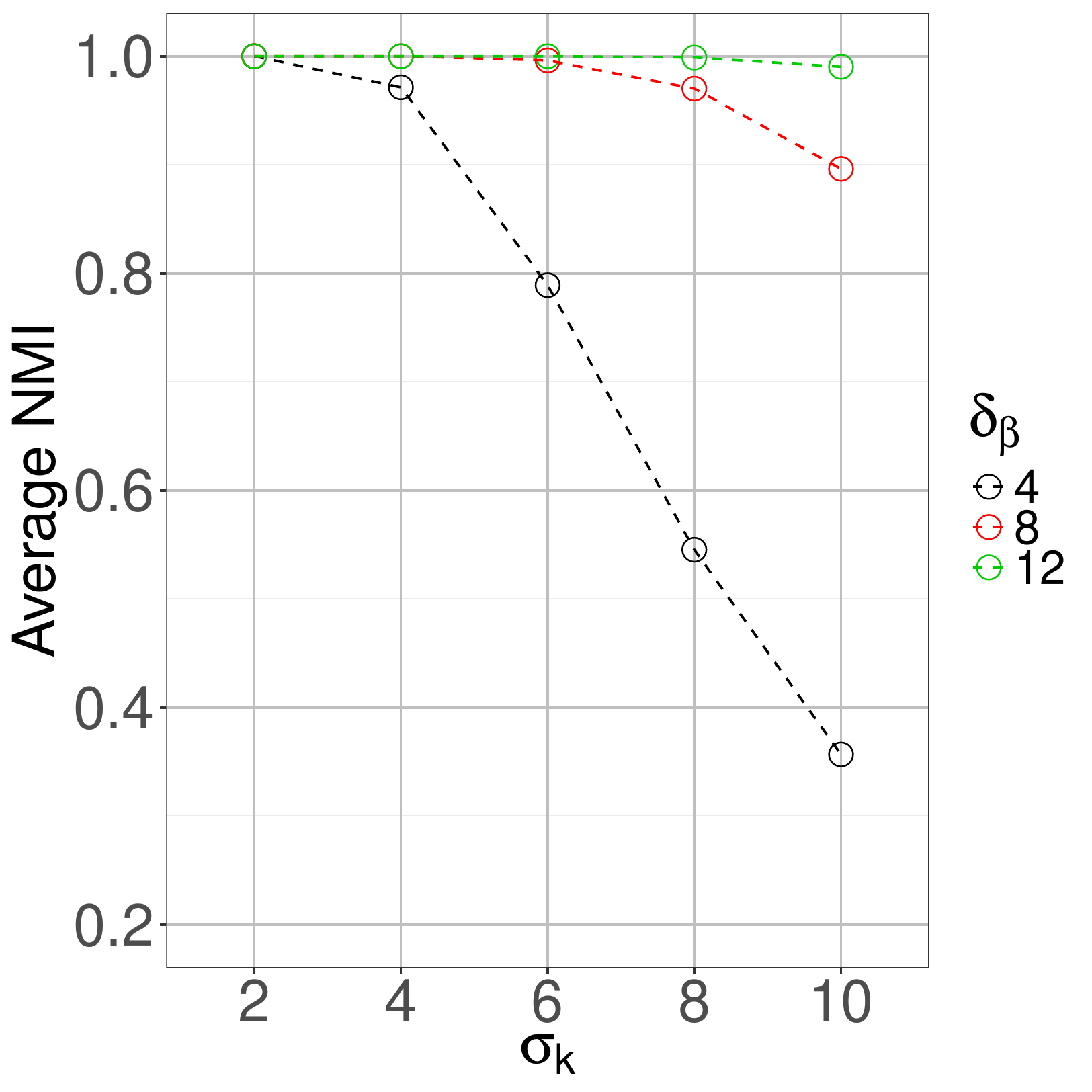}
      \caption{\footnotesize{}}\label{sfig:nmi_pk2}
    \end{subfigure}
    \begin{subfigure}[b]{0.32\textwidth}
      \centering
      \includegraphics[width=\textwidth]{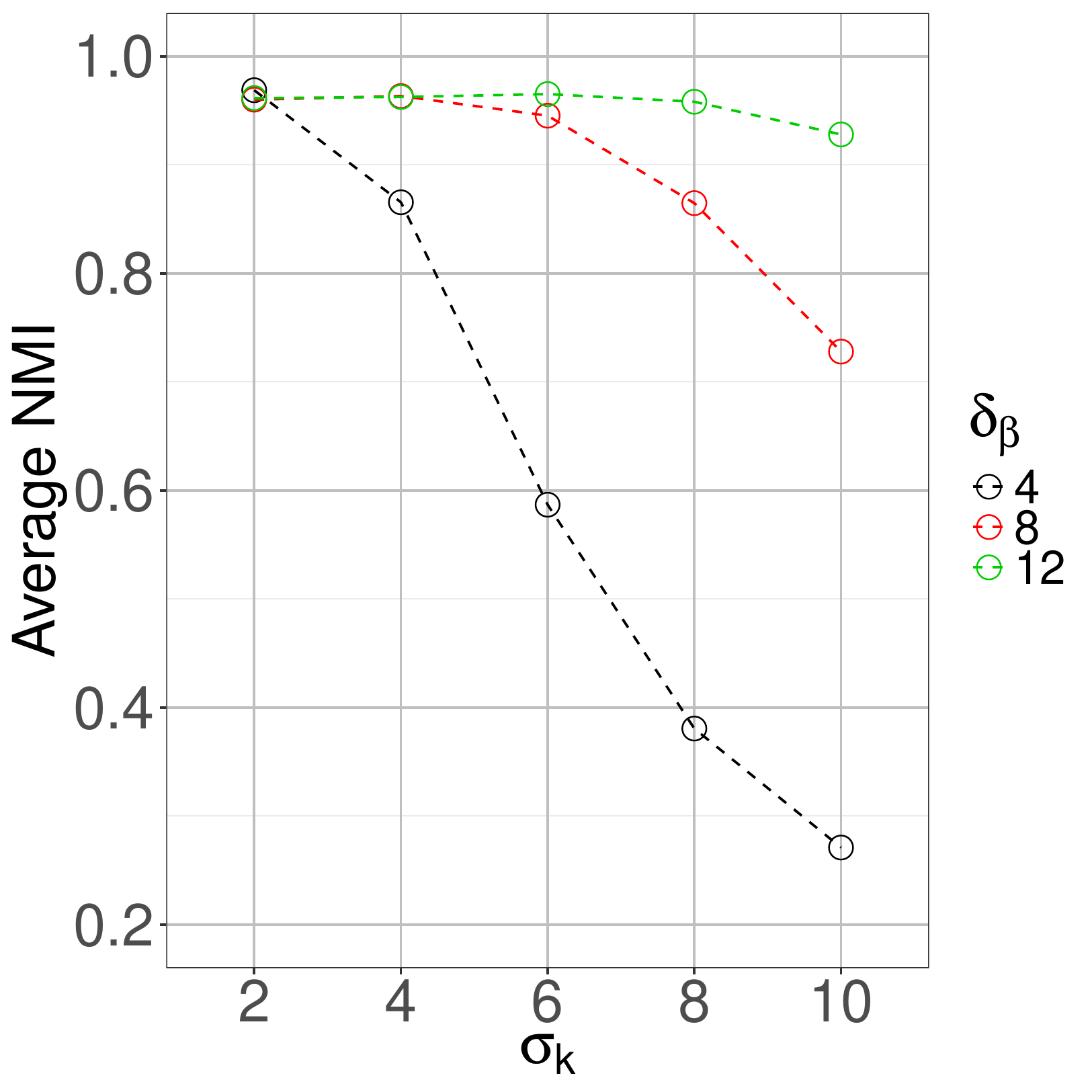}
      \caption{\footnotesize{}}\label{sfig:nmi_pk3}
    \end{subfigure}
    \caption[The impact of $K$ and $p$ on NMI for the case $n=400$]{The impact of $K$ and $p$ on NMI for the case $n=400$; each colored line in a plot represents different value of $\delta_\beta$, $x$ axis shows different values of $\sigma_k$, and $y$ axis is average NMI for: (a) $K=2$, $p=2$, (b) $K= 2$, $p=4$, (c) $K=4$, $p=4$.
    }
  \end{figure}

\paragraph{Dimensionality ($p$) and Number of Clusters ($K$).}
Figure \ref{fig:p:k} shows the NMI result for different combinations of $p$ and $K$, with $n=400$. Figure \ref{sfig:nmi_pk1} and \ref{sfig:nmi_pk2}, compare the result when $K$ is fixed ($K=2$) and the dimensionality is changed from $p=2$ in \ref{sfig:nmi_pk1} to $p=4$ in \ref{sfig:nmi_pk2}. Comparing the two plots, we can see slight improvement in the case where $p=4$. To study the effect of increasing $K$, we can compare Figures \ref{sfig:nmi_pk2} and \ref{sfig:nmi_pk3}, where $p$ is fixed ($p=4$), and $K$ is increased from $K=2$ (Figure \ref{sfig:nmi_pk2}) to $K=4$ (Figure \ref{sfig:nmi_pk3}). We can clearly see that the accuracy decreases in all cases (combinations of $\delta_\beta$ and $\sigma_k$). This is consistent with the fact that as the number of clusters ($K$) increases, it is always harder to recover true clusters. The results for $\beta$-estimation error are reported in Figure~\ref{fig:p:k:bet} in Appendix~\ref{sec:tables}.

\label{fig:p:k}

\paragraph{Number of Groups in a Cluster.} 
To study the impact of the number of observations per group, we set the total number of observations to $n=100$ and take $K=2$, which results in each cluster having 50 observations. We then vary the number of groups per cluster ($G$) from 50 down to 1 for each cluster, with $G=50$ referring to the case where each observation is a single group, i.e., there is no grouping structure; whereas $G=1$ is the case where all the observations in each cluster form a single group. In this setup, when varying $G$, we do not necessarily preserve equal number of observations per group ($n_r$), as in other experiments. If the observations can be equally distributed to all groups i.e. $G\in{\{50, 25, 10, 5, 2, 1\}}$, then there will be equal number of observations per group. In other cases, the observations are first equally distributed among the groups. The remaining observations are then assigned to the groups in a fashion that a group gets one extra observation until all the observations are distributed. 
For example, when $G=48$, there will be 2 groups with 2 observations and 46 groups with single observations per cluster. Similarly, when $G =16$, there will be 2 groups with 4 observations and 14 groups with 3 observations per cluster. 

\begin{figure}
 	\centering
 	\includegraphics[width=.6\textwidth]{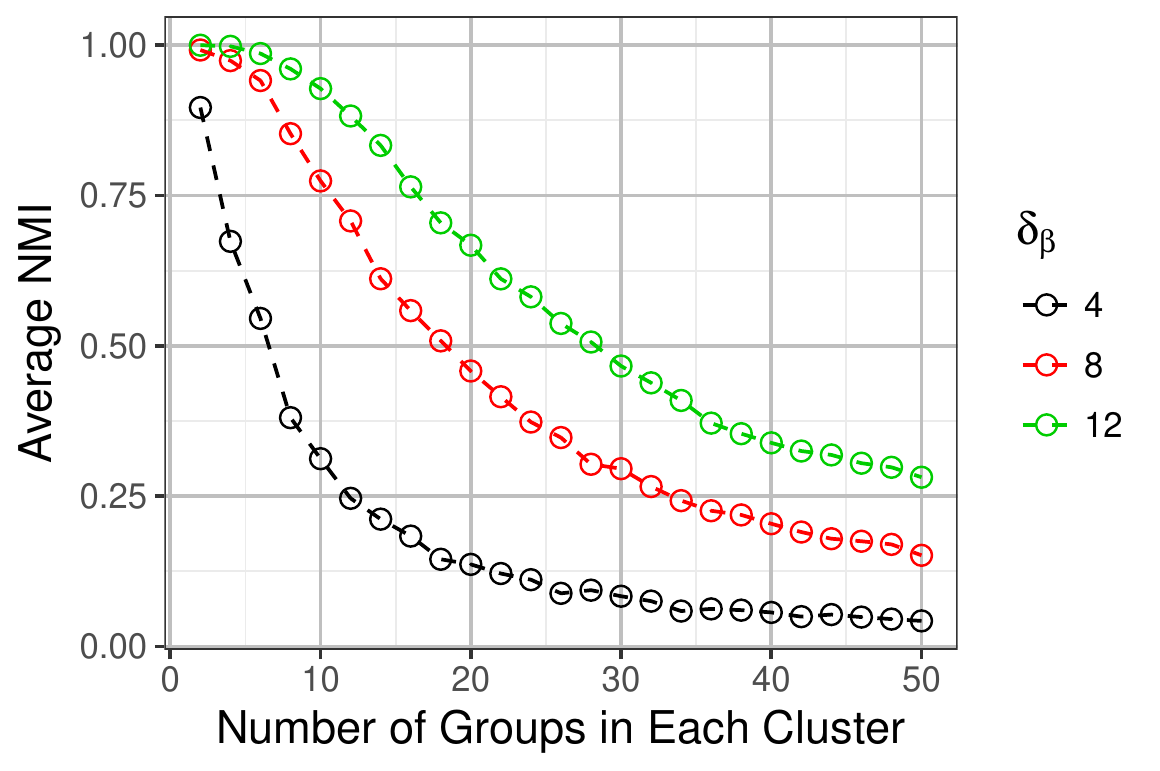}
 	\caption[The effect of the number of groups per cluster ($G$) for the case $n=100, K=2$]{The effect of the number of groups per cluster ($G$) for the case $n=100, K=2$; each colored line in a plot represents a different value of $\delta_\beta$; $x$ axis shows $G$ which varies from 1 to 50 (only even numbers are plotted to enhance the quality). $y$ axis shows the average NMI.}
	\label{fig:R_perform}
 \end{figure}

Figure \ref{fig:R_perform} illustrates the result. We notice that as $G$ increases, average NMI decreases, indicating the difficulty to recover the true class labels. This is expected because as we have more groups ($G$ is larger), there are less observations in each group, which makes it difficult to utilize the grouping structure and therefore the performance drops.

%

\paragraph{Number of Iterations.} 
One of the important factors in determining the effectiveness of an algorithm is its rate of convergence and its overall run time, which determines its suitability for high dimensional applications. Since the run time depends on several factors such as the platform, the quality of coding, hardware, etc. it is hard to report an accurate value for an algorithm. We report the average number of iterations for GMR convergence in each scenario as an estimated indicator for the speed of the algorithm. 
 
\begin{figure}
    \centering
    \begin{subfigure}[b]{0.49\textwidth}
      \centering
      \includegraphics[width=\textwidth]{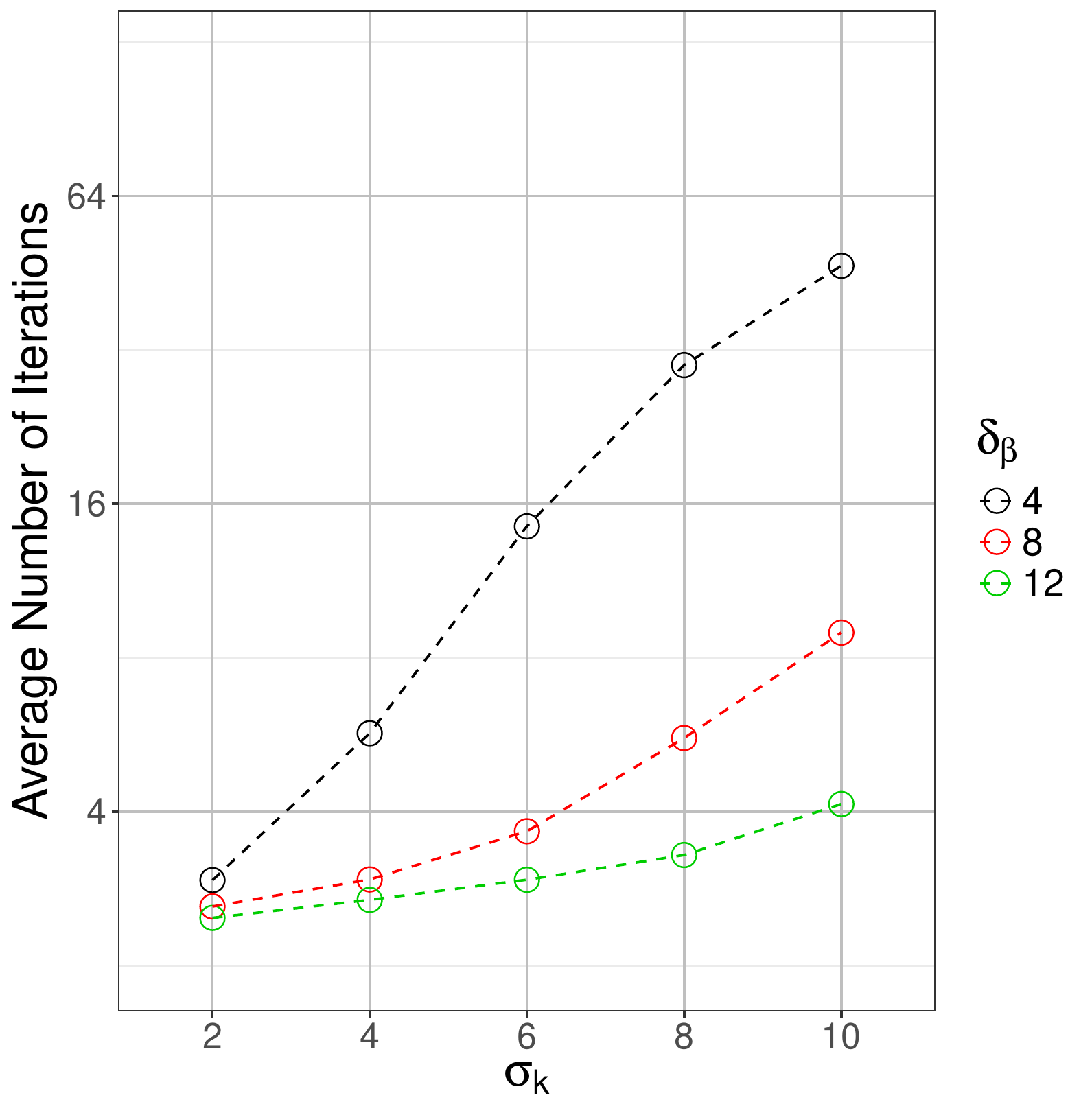}
      \caption{\footnotesize{}}\label{sfig:itr1}
    \end{subfigure}
    \begin{subfigure}[b]{0.49\textwidth}
      \centering
      \includegraphics[width=\textwidth]{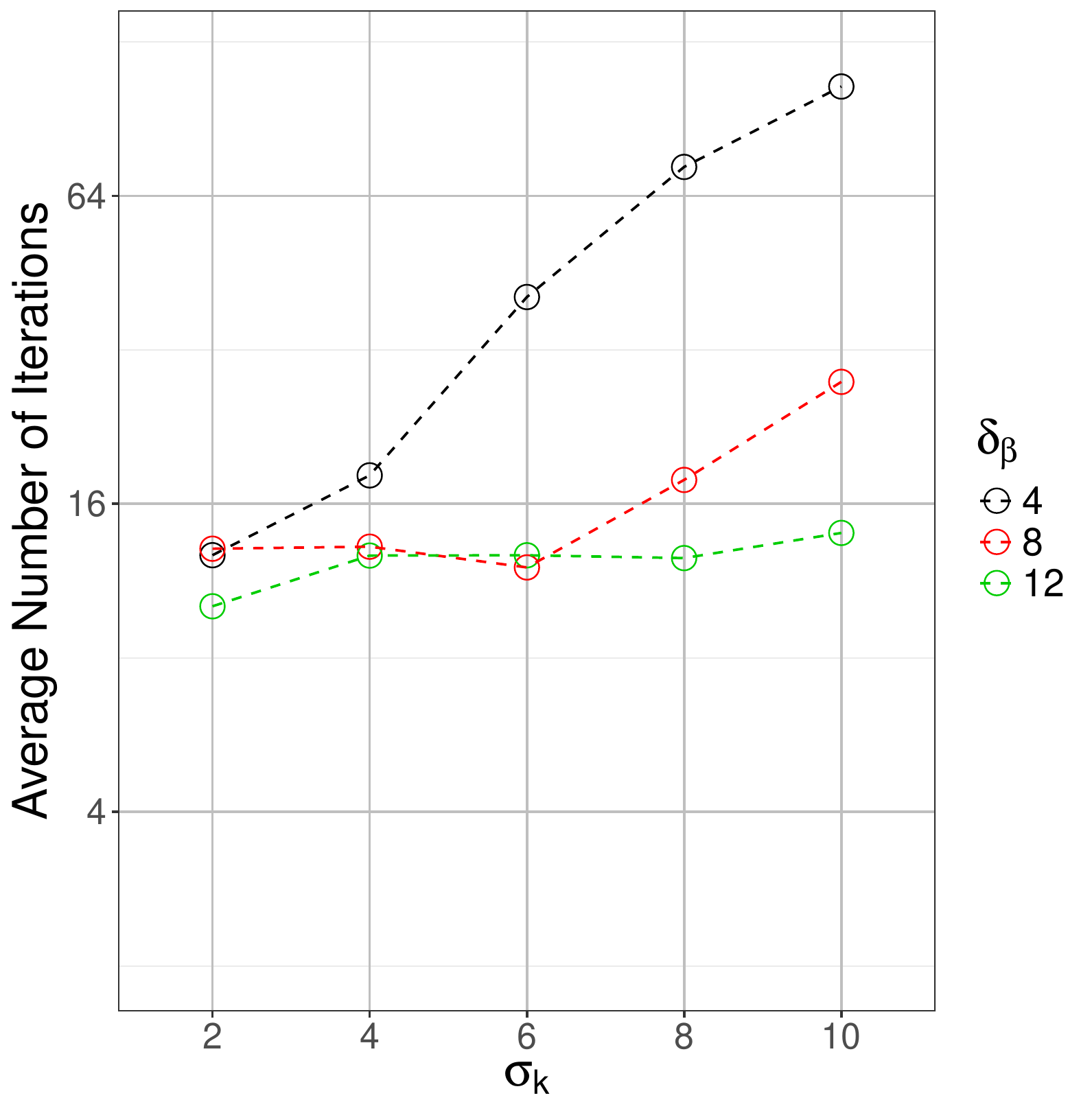}
      \caption{\footnotesize{}}\label{sfig:itr3}
    \end{subfigure}
    \caption[Average number of iterations for the case $n=800$]{Average number of iterations for the case $n=800$ (on $log_2$ scale) for: (a) $K=2$, $p=2$, (b) $K=4$, $p=4$.
  	}
  \label{fig:itr}
\end{figure}

The stopping rule is chosen to be the relative change in posterior probability of cluster assignments ($\tau_{rk}(\hat{\theta})$ in equation \eqref{eq:posterior:memebership}). In particular, if we call $\tau_{rk}^{(t)}$ the posterior probabilities at iteration $t$, then the algorithm stops when $\|\tau^{(t-1)}-\tau^{(t)}\|_\infty<\epsilon$, where $\norm{\cdot}_\infty$ is the infinity norm (maximum absolute row sum), or if the maximum number of iterations has been reached. In our setup, $\epsilon$ is set to $10^{-6}$ and the maximum number of iterations to 200. Figure \ref{fig:itr} shows average number of iterations for selected scenarios. The $y$ axis is shown in $log_2$ scale to enhance visualization. Tables (\ref{tab:nmi}--\ref{tab:itr}) in Appendix~\ref{sec:tables} provide full details of the results for the conducted simulation and modeling experiments mentioned in this section.

\subsection{Selecting Optimal $K$}\label{subsec:opt:k}
Selecting the number of components in a mixture model, which falls under the general problem of model selection, is a research topic that has attracted a lot of interest over years. Despite numerous advances, the problem is a fairly open question in statistics and machine learning, at least from a practical standpoint. Among the numerous methods introduced in the literature for determining the optimal number of clusters in a dataset, we refer to the following selective samples: \cite{goutte1999clustering}, \cite{pelleg2000x}, \cite{goutte2001feature}, \cite{lleti2004selecting} and~\cite{honarkhah2010stochastic}. 

In the presence of independent variable(s), Cross-Validation (CV) is a simple and popular way of selecting the tuning parameters, including the best choice of $K$. To investigate the performance of CV under GMR in determining $K$, we set up an experiment with $\delta_\beta=(8,12), \sigma_k=6, \text{and } N=200$, where the data was generated using a true $K^*=4$. GMR is trained on the data using $K=2,\dots, 8$ clusters, and in each case the trained model is used to predict a held out set. 

\begin{figure} 
\centering
\includegraphics[width=3.2in]{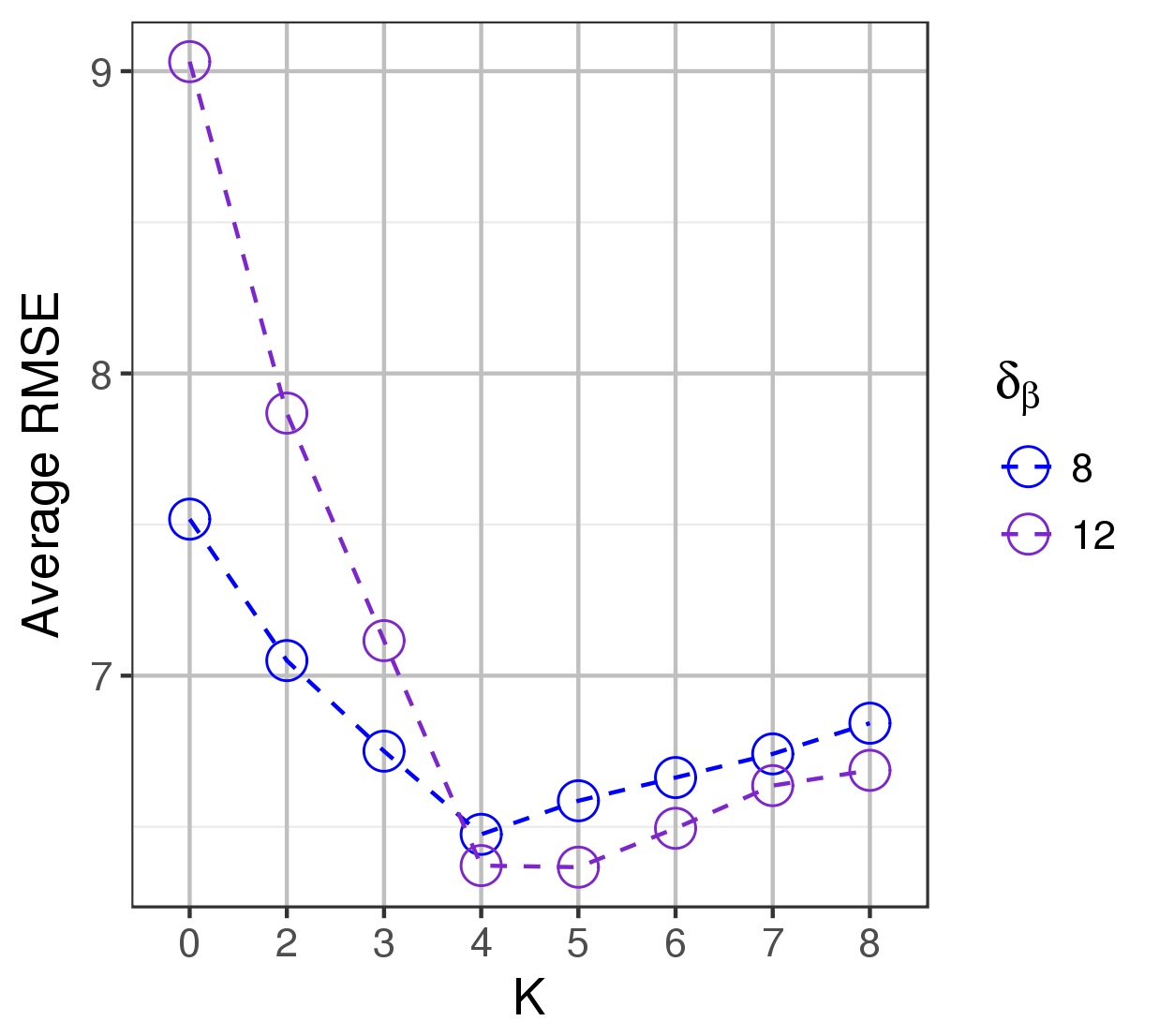}
\caption[Finding the optimal value of $K$ using cross validation]{Finding the optimal value of $K$ using cross validation: true $K^*=4$. GMR is applied with different numbers of $K \in {\{0,\dots,8\}}$ and they are shown in $x$ axis in the graph. $K=0$ refers to prediction by mean while $K=1$ is the result of prediction using a single linear regression model.}
\label{fig:k_perform}
\end{figure}

Figure \ref{fig:k_perform} shows the plot of the average prediction RMSE against the $K$ used in training.
To compare the performance with some baselines, the test data is predicted using the mean (response $y$) of the training data as well as a single linear regression model. These two cases correspond respectively to $K=0$ and $K=1$ in Figure \ref{fig:k_perform}. One clearly observes that the GMR with $K > 1$ outperforms both the mean prediction ($K=0$) and a single linear model ($K=1$). The minimum error among the GMR models is attained with $K = 4$, which is the true number of components. The plot validates the ability of the algorithm to find the optimal number of components using CV. 


\subsection{Prediction Performance}
\begin{figure} 
\centering
\includegraphics[scale=0.5]{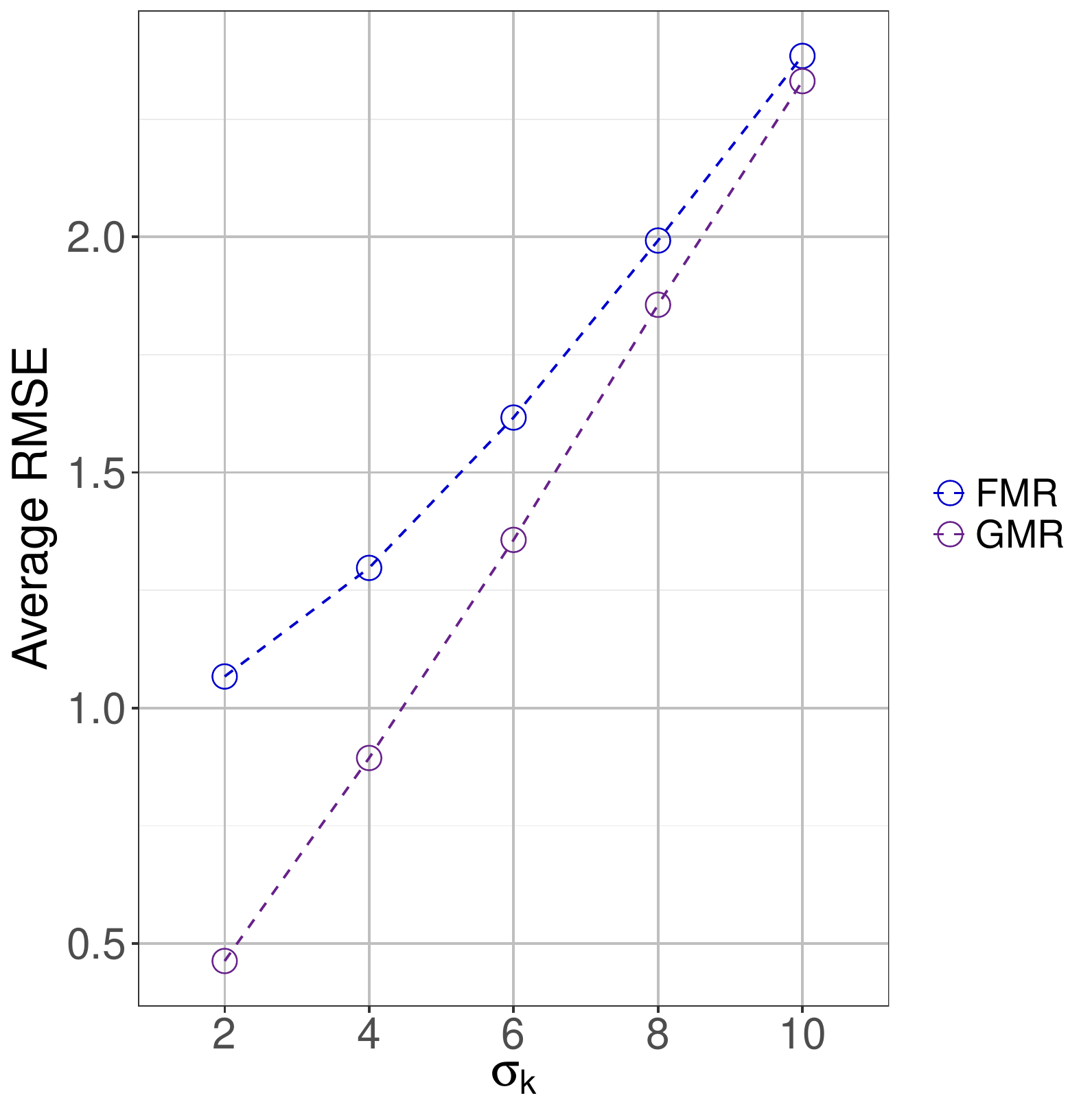}
\caption[The MAP prediction accuracy of the GMR versus the regular FMR]{MAP prediction accuracy of the GMR (purple) versus the regular FMR (blue).}
\label{fig:predict:compare}
\end{figure}

As noted earlier in Section~\ref{sec:post:predict}, the advantage of GMR over regular FMR is the posterior predictive density that enables us to utilize prior information about the group that a new observation is coming from. We claim that utilizing this prior knowledge can lead to a better prediction accuracy. To test the robustness of the prediction power of the model, we sample data by setting $n=200, K=p=4, \delta_\beta=8, G=10$ and train the GMR using the training data in each group. We then predict the hold-out set, first with the regular FMR and then with GMR. Note that in the case of regular FMR, once the model is trained and the parameters of the models are estimated, new observations will be predicted using the standard mixture model rule. In our group structure setup, we obtain the parameter estimates $\thh =(\beh_k, \sigh_k, \pi_k;k\in{\{1,\dots,K\}})$ in the training phase. The MAP prediction using regular FMR is $y_{\text{new}} = \sum_{k=1}^K \pi_k \beh_k^T x_{\text{new}}$ while that of the GMR is $\ynew{r} = \sum_{k=1}^K \tau_{rk}(\thh) \,\phi_{\sigma_k}\big( \ynew{r} - \beh_k^T \xnew{r} \big)$ as discussed in Section \ref{sec:post:predict}.
Figure~\ref{fig:predict:compare} shows the resulting average RMSEs versus $\sigma_k$. It is clear that GMR prediction outperforms that of FMR indicating the improvement brought about by incorporating group membership information.

\subsection{Comparing GMR with \mmclpp}
\begin{figure} 
    \centering
    \begin{subfigure}[b]{0.49\textwidth}
      \centering
      \includegraphics[width=\textwidth]{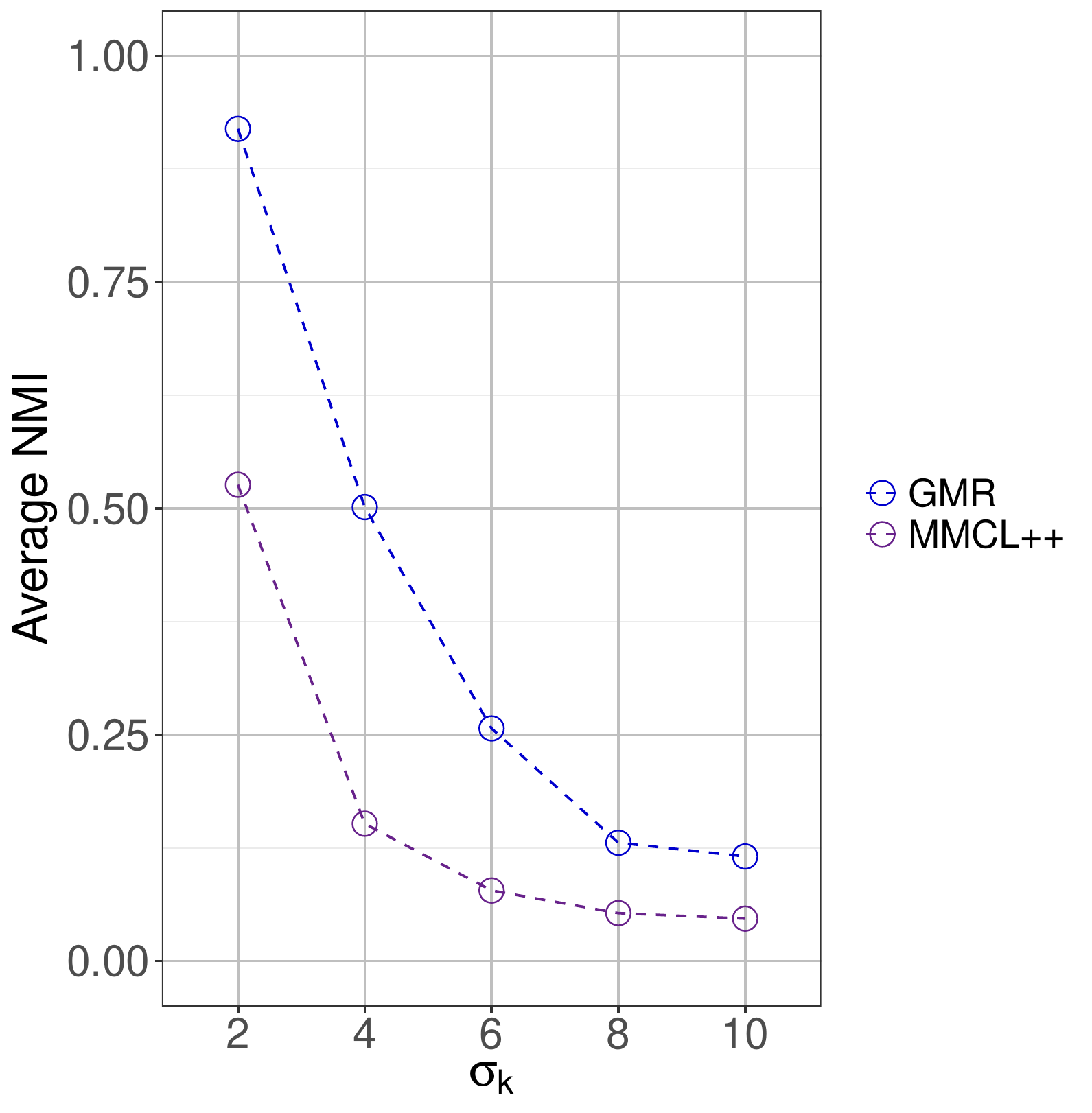}
      \caption{\footnotesize{}}\label{sfig:MMCL:GMR:bet4}
    \end{subfigure}
    \begin{subfigure}[b]{0.49\textwidth}
      \centering
      \includegraphics[width=\textwidth]{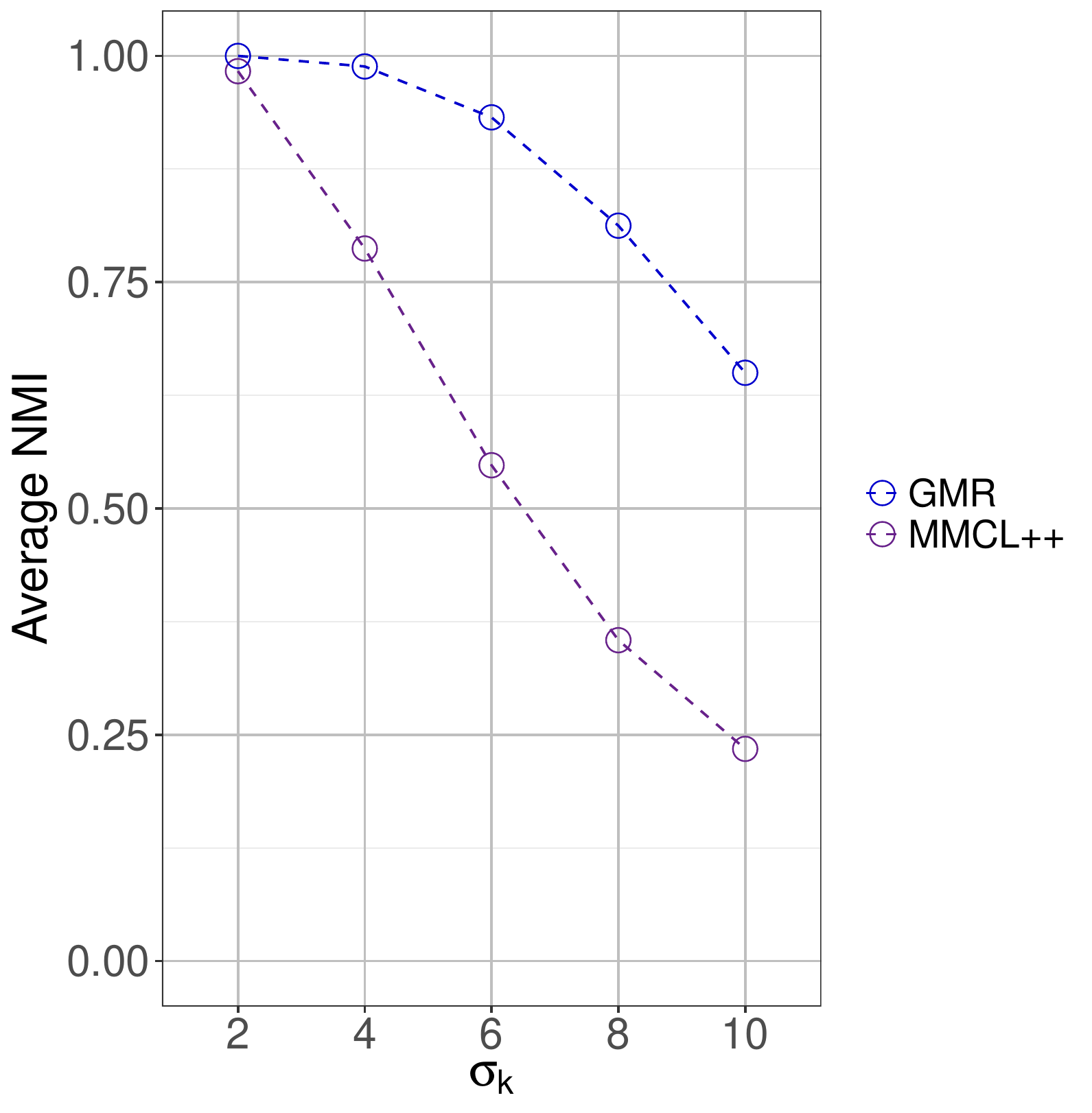}
      \caption{\footnotesize{}}\label{sfig:MMCL:GMR:bet12}
    \end{subfigure}
    \caption[Label recovery of the \mmclpp versus the GMR: Average NMIs]{Label recovery of the \mmclpp versus the GMR: Average NMIs for (a) $\delta_\beta=4$, (b) $\delta_\beta=12$. In all cases $n=100$.}
\label{fig:MMCL:GMR}
\end{figure}

\begin{figure} 
    \centering
    \begin{subfigure}[b]{0.49\textwidth}
      \centering
      \includegraphics[width=\textwidth]{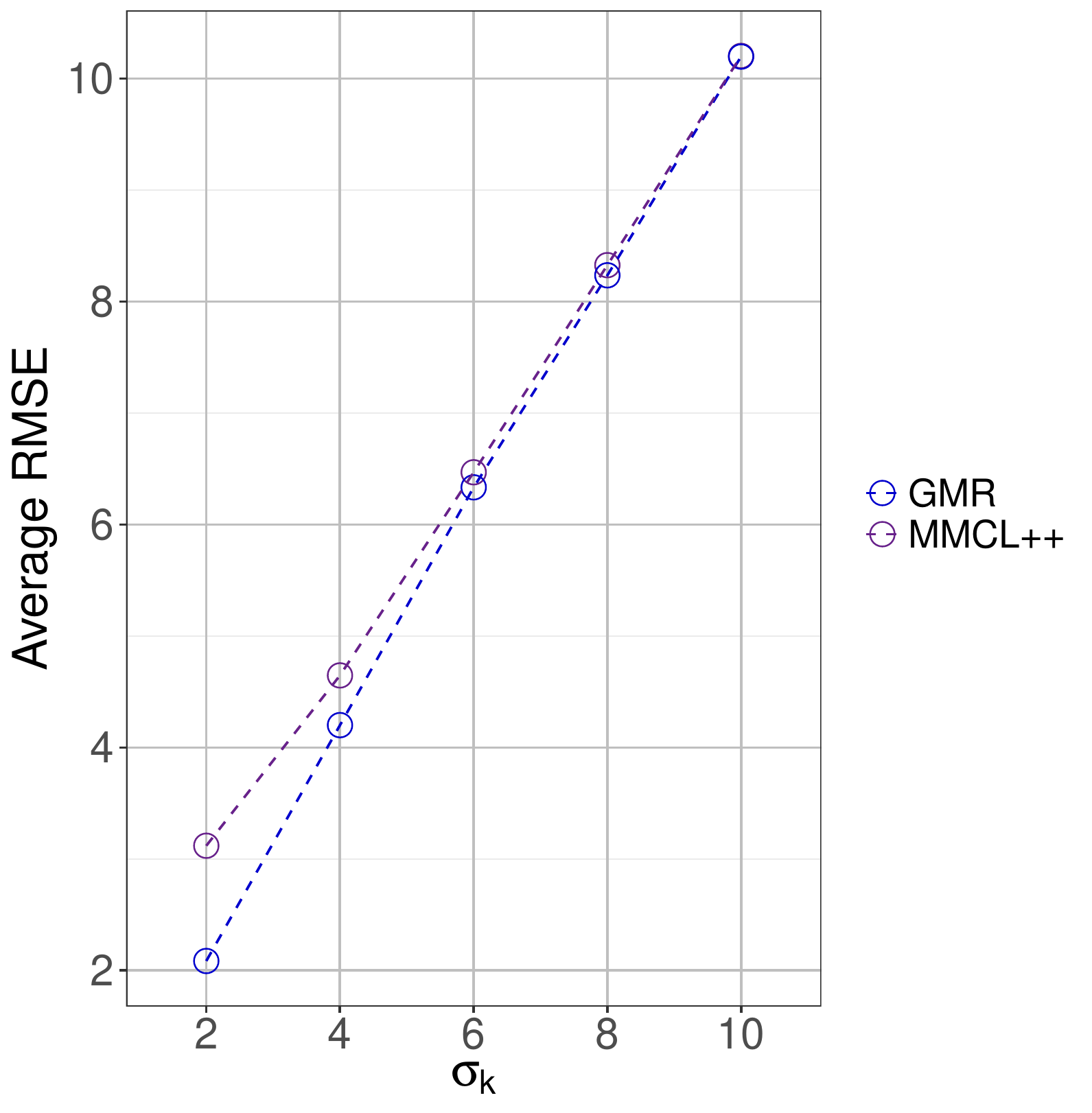}
      \caption{\footnotesize{}}\label{sfig:rmse:MMCL:GMR:bet4}
    \end{subfigure}
    \begin{subfigure}[b]{0.49\textwidth}
      \centering
      \includegraphics[width=\textwidth]{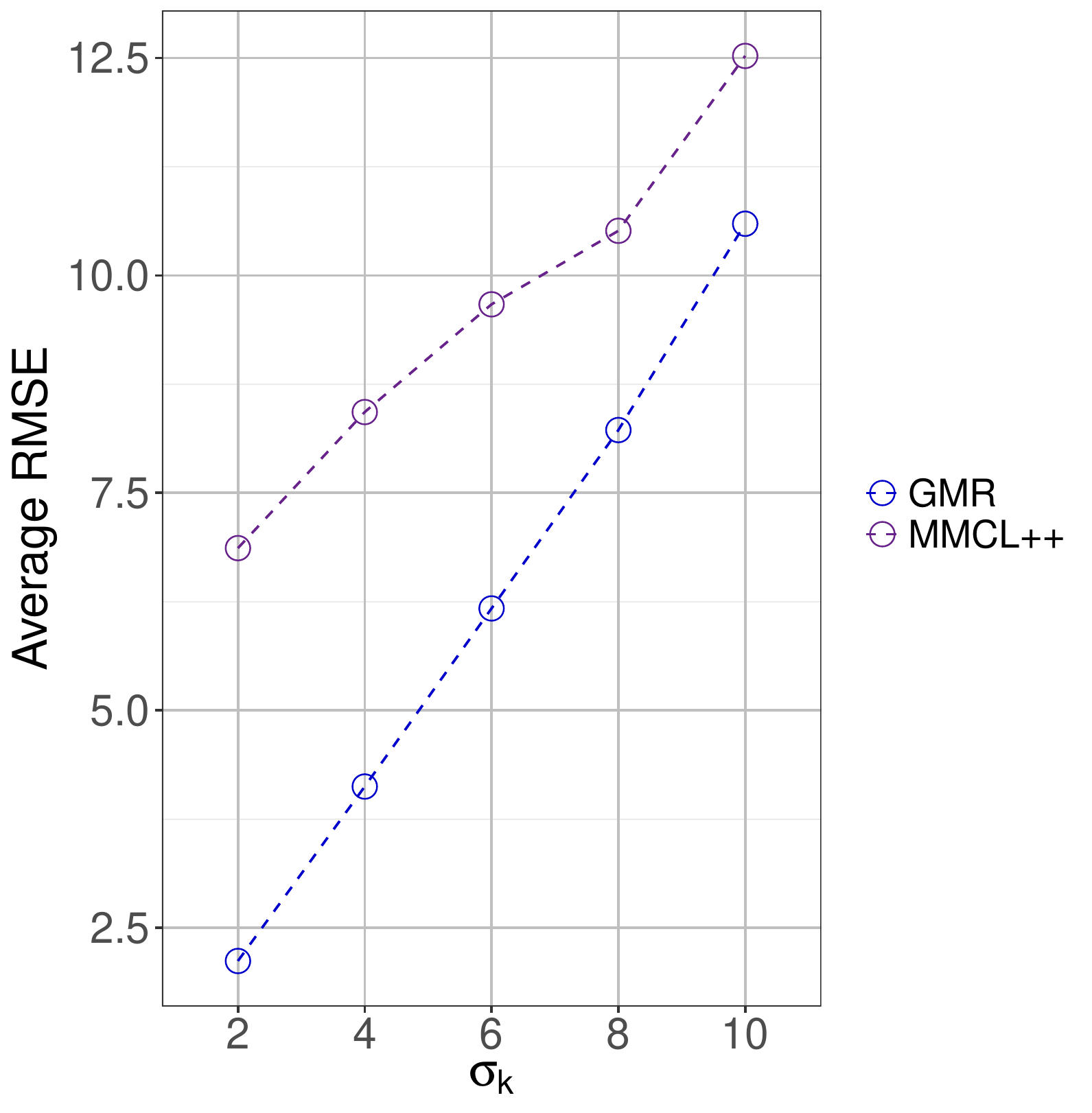}
      \caption{\footnotesize{}}\label{sfig:rmse:MMCL:GMR:bet12}
    \end{subfigure}
    \caption[Comparing the prediction power of \mmclpp and GMR]{Comparing the prediction power of \mmclpp and GMR: Average RMSE value ($n=100$) for (a) $\delta_\beta=4$, (b) $\delta_\beta=12$}
\label{fig:rmse:MMCL:GMR}
\end{figure}

To perform a comparison between GMR and the existing method \mmclpp~\citep{haidar:inpress}, we ran both algorithms on the same data generated according to Table~\ref{tab:exp}. Figure~\ref{fig:MMCL:GMR} illustrates the comparison of their ability for label recovery for the case $n=100$ and $G=10$. We can see that GMR outperforms \mmclpp in all cases in terms of correctly recovering the true labels. Figure \ref{fig:rmse:MMCL:GMR} compares the prediction power of the two algorithms in terms of the average RMSE; the prediction setup is as described earlier. Although both algorithms are very close in prediction performance, at higher uncertainties ($\sigma_k >5 \text{ and } \delta_\beta=4$), GMR outperforms \mmclpp in predicting new observations.

\section{Case Study: Dealership Performance Assessment}
In this section, we present the results from applying the proposed GMR to a real-world problem in the retail industry. We show how to use GMR to provide guidelines and recommendations for improving the performance of retail stores, and in particular, the automotive dealerships. We apply the GMR to fit mixture of regression models to a dealership dataset in order to cluster the stores while accounting for the similarities in store performance dynamics. 

\subsection{Dealership Dataset}
\begin{figure} [t]
\centering
\includegraphics[scale=0.7]{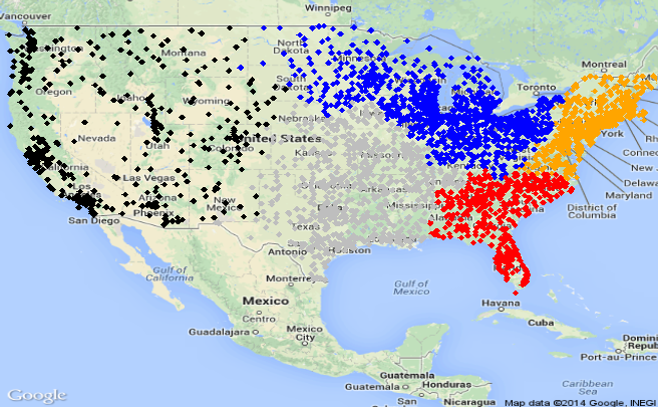}
\caption[Network of OEM dealerships in the U.S.]{Network of OEM dealerships in the continental U.S. grouped by regions: Northeast (golden), Southeast (red), Great Lakes (blue), Central (gray), and West (black)}
\label{fig:map}
\end{figure}

For reasons of confidentiality, we are not able to reveal full details about the dataset. The dataset made available to us consists of several thousands (3,074) dealerships, with consecutive monthly financial data (observations) for each dealer spanning five years (60 observations per dealer). Figure~\ref{fig:map} shows the dealer network across the United States. It shows how the dealers are distributed and grouped into five regions: Northeast (yellow), Southeast (red), Great Lakes (blue), Central (gray), and West (black). There were 281 Key Performance Indicators (KPIs) in the monthly financial documents deemed important by the domain experts. We treat these KPIs as independent variables and standardize them to have zero mean and unit standard deviation.

To prepare the data for the application of the GMR, the observations for all the dealers are first aggregated to construct the design matrix $X \in \reals^{(3074\times60)\times281}$. Since the data for each dealership is generated for each month, we checked for trends and seasonality for each dealer and found no evidence that there exists trends or seasonality between the consecutive months. The reason is that the KPIs are constructed in a way that the trend and seasonality are absorbed by a special normalization procedure. The observations for each dealer are treated as a single group and assigned a unique group ID. That is, the number of groups in our setup is the same as the number of dealers.

\subsection{Dealership Performance Prediction} 
In this section, we provide the results from applying the proposed GMR approach for modeling the productivity of automotive dealerships across the U.S. for a particular Original Equipment Manufacturer (OEM). We compare the results of the GMR with those of the \mmclpp \citep{haidar:inpress}. Because of the large size of the dataset and specially the large number of predictors, Least Absolute Shrinkage and Selection Operator (LASSO) technique \citep{tibshirani1996regression} is used for regression modeling of both the sales as well as the profitability when \mmclpp is applied. To apply the GMR, the number of clusters ($K$) has to be identified in advance. As demonstrated by the simulations in Section~\ref{subsec:opt:k}, GMR can properly select the true $K$ by cross-validation. The same process is applied to the dealership dataset to find the best $K$.

\subsubsection{Results}

\begin{figure}[t]
\centering
\begin{subfigure}{.49\textwidth}
\centering
\includegraphics[width=0.9\linewidth]{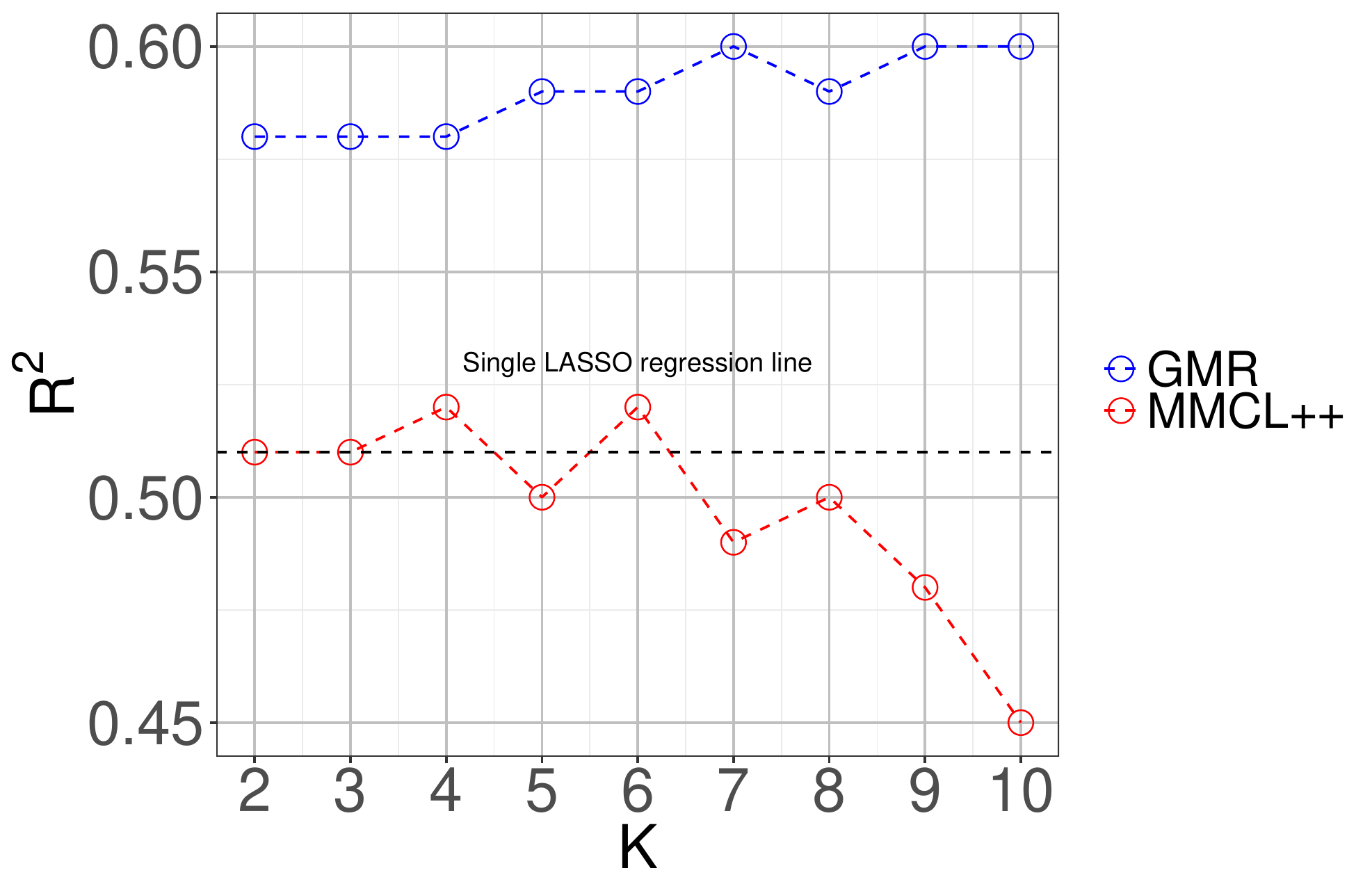}
\caption{\footnotesize{}}
\label{sfig:r2:ROS}
\end{subfigure}%
\begin{subfigure}{.49\textwidth}
\centering
\includegraphics[width=0.9\linewidth]{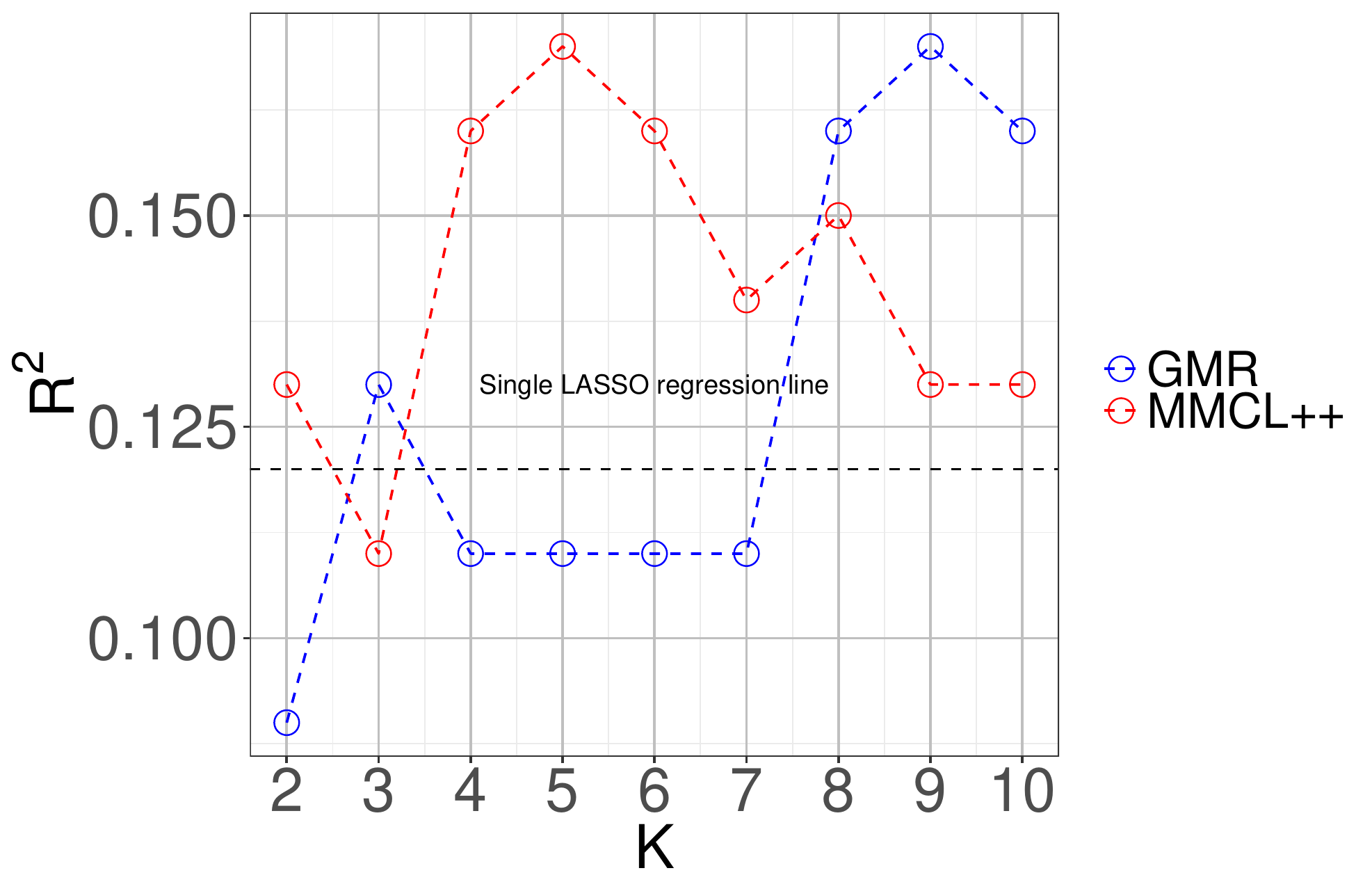}
\caption{\footnotesize{}}
\label{sfig:r2:SE}
\end{subfigure}
\caption[Results from applying \mmclpp and GMR to the dealership dataset ]{Results from applying \mmclpp, and GMR to the dealership dataset with two dependent variables: (a) profitability and (b) sales effectiveness. The horizontal black dashed line is the $R^2$ for a single LASSO model.}
\label{fig:r2:res}
\end{figure}

Figure~\ref{fig:r2:res} summarizes the results. The plots report the prediction $R^2$ (i.e., over the test portion of the dataset) for both the performance metrics (``profitability'', the concern of the dealership, and the ``sales effectiveness'', the OEM's main objective). We are displaying the results from running the GMR and \mmclpp for $K \in \{2,\dots,10\}$. The case $K=1$ (the horizontal dashed line in the figure) corresponds to fitting a single linear regression model to the entire training dataset.

As the results suggest, GMR has improved the accuracy for predicting both the profitability and sales effectiveness metrics. It was able to achieve a $R^2$ value of 0.6 using $K=9$ and $K=10$, whereas the model with a single component has $R^2$ of 0.51, a 9\% improvement. The highest $R^2$ that \mmclpp was able to achieve for profitability is 0.52 (with $K=4$ and $K=6$). In the case of sales effectiveness, the single component model is able to produce a $R^2$ value of 0.12. However, GMR was able to improve this value to 0.17 (41\% improvement) with $K=9$. \mmclpp also produced the same result with $K=5$. This result also suggests that there is heterogeneity among the dealers and by clustering them, one can improve the analysis and generate better recommendations to dealers for improving their performance.

Reviewing Figure~\ref{fig:r2:res}, we conclude that if GMR is used, we should ideally cluster the dealers into 9 groups where the models show the highest $R^2$ for both the profitability and the sales effectiveness. In the case of modeling with \mmclpp, it is best to partition the dealers into 4 clusters. 

It should be mentioned that in large datasets such as our dealership problem, the \mmclpp approach is computationally more expensive. This is in general true when the number of groups is large, because the algorithm has to extract, model, and evaluate the results of all the groups in every iteration. The issue is even more problematic when employing cross-validation to find the best tunning parameters (such as $K$) as well as establishing the initial groups.
This is not the case for GMR, for it had no problem handling and producing the results for such a large dataset in seconds on a regular laptop.

\vspace{3mm}

\subsubsection{Assessing the Clusters}
\begin{figure}[ht]
\begin{subfigure}{.49\textwidth}
 \centering
 \includegraphics[width=1\linewidth]{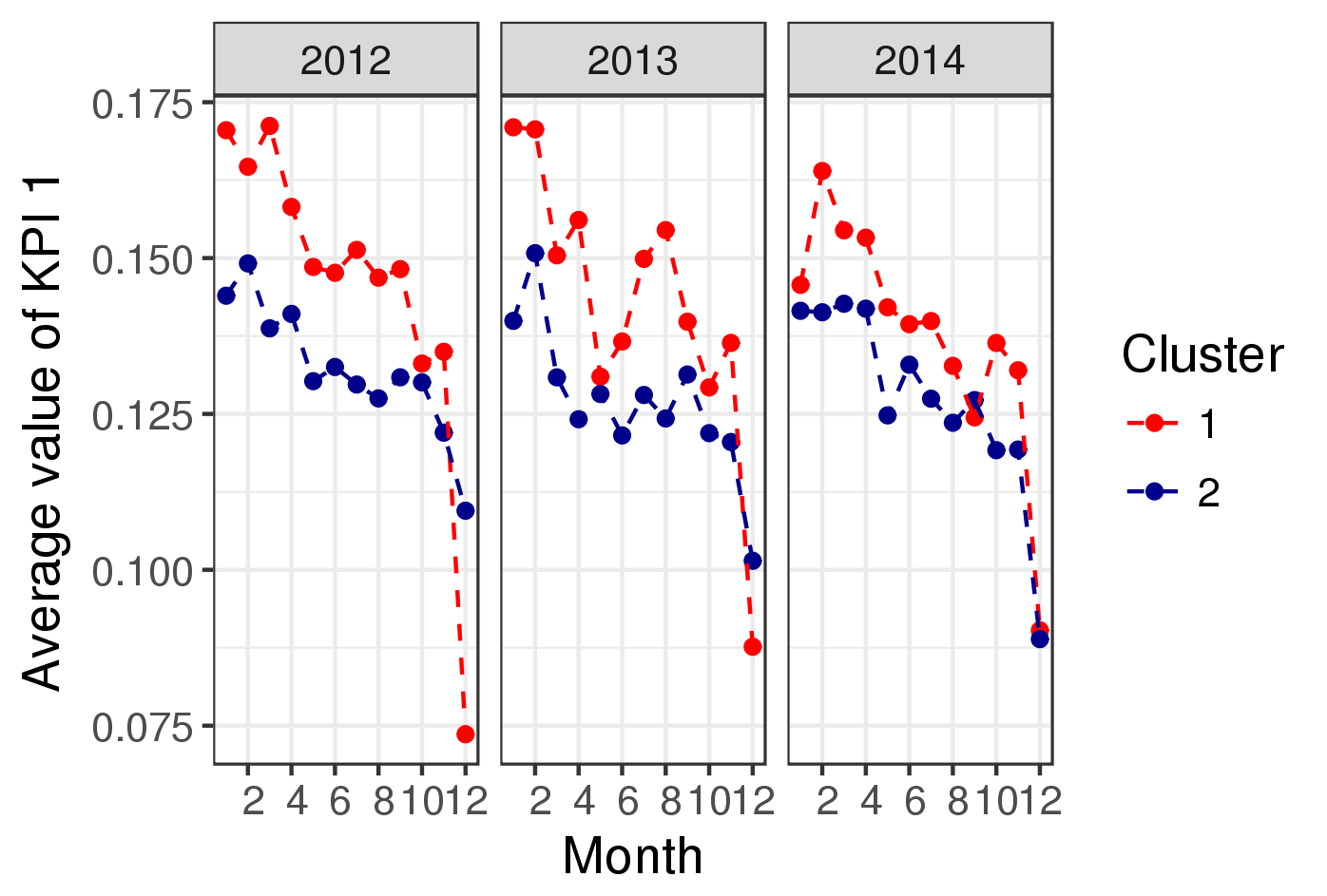}
 \caption{\footnotesize{}}
 \label{sfig:clust:eval1}
\end{subfigure}%
\begin{subfigure}{.49\textwidth}
 \centering
 \includegraphics[width=1\linewidth]{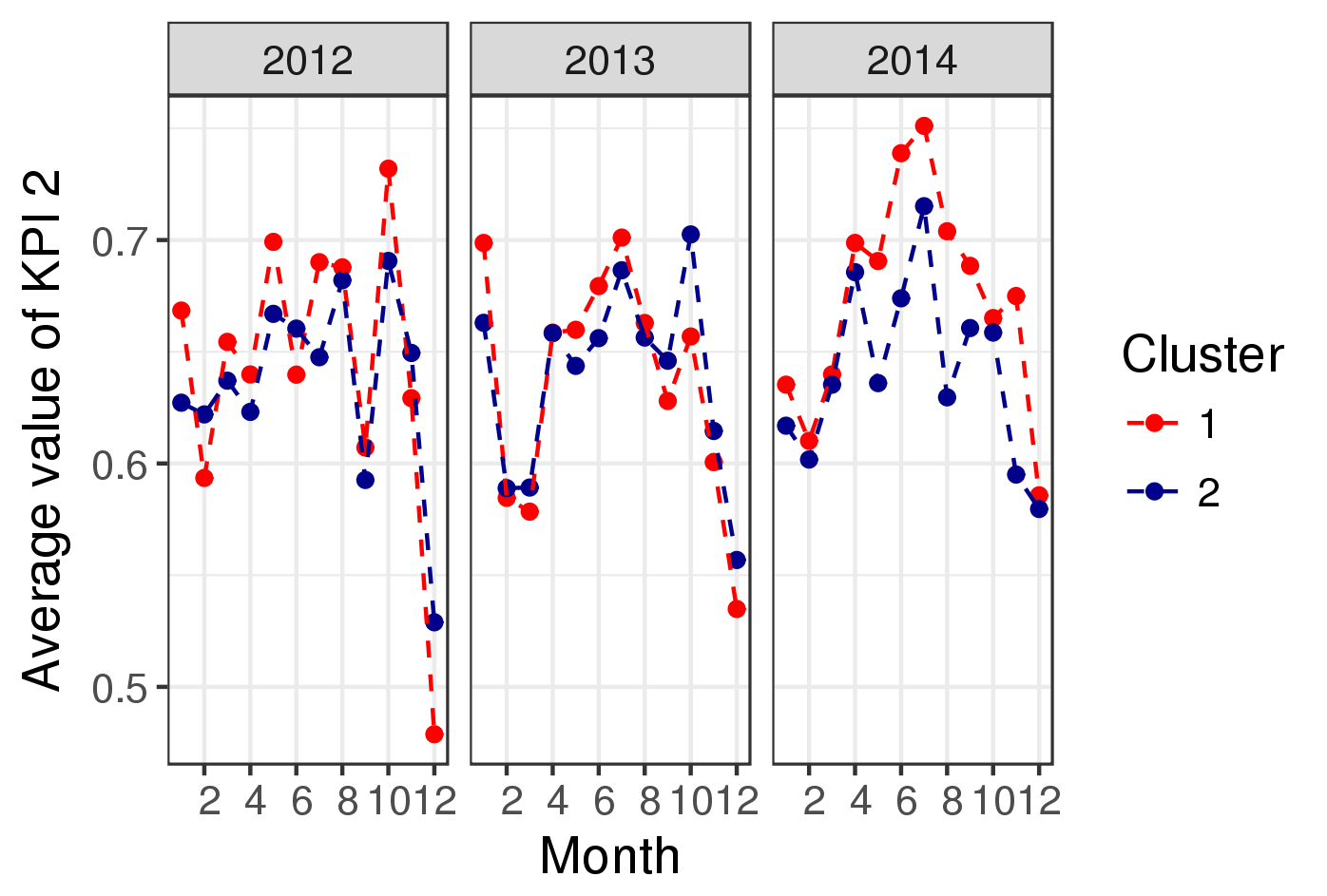}
 \caption{\footnotesize{}}
 \label{sfig:clust:eval2}
\end{subfigure}
\caption[Assessing the Clusters formed by GMR in the KPI Space]{Assessing the clusters formed by GMR in the KPI space. The average value for each month and year is displayed for Cluster 1 (red) and Cluster 2 (blue) (a) KPI \#1 (b) KPI \#2}
\label{fig:clust:eval}
\end{figure}

To evaluate the clusters resulting from GMR, we applied the GMR to a region within the U.S. as requested by the domain experts. The number of clusters $K$ is set to 2. The following plots are produced to visually evaluate the effectiveness of the formed clusters. Figure~\ref{fig:clust:eval} shows the average values for the two most important KPIs (i.e., those with the highest regression coefficients), plotted for the two clusters: cluster~1 in red and cluster~2 in blue. The plots only includes the last 36 months of the data for a better visualization.

\begin{figure}[t]
\begin{subfigure}{.5\textwidth}
 \centering
 \includegraphics[width=1\linewidth]{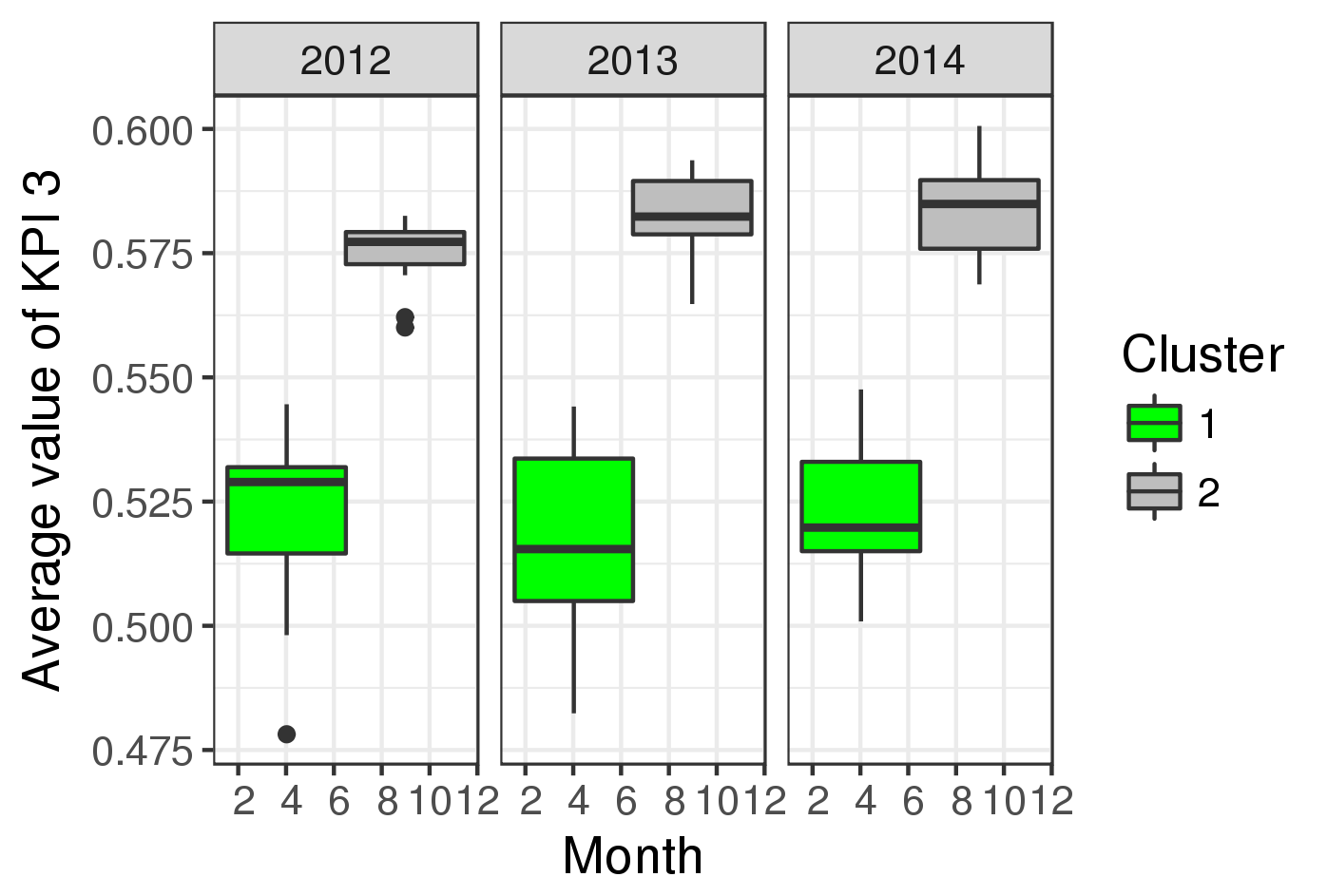}
 \caption{}
 \label{sfig:clust:box1}
\end{subfigure}%
\begin{subfigure}{.5\textwidth}
 \centering
 \includegraphics[width=1\linewidth]{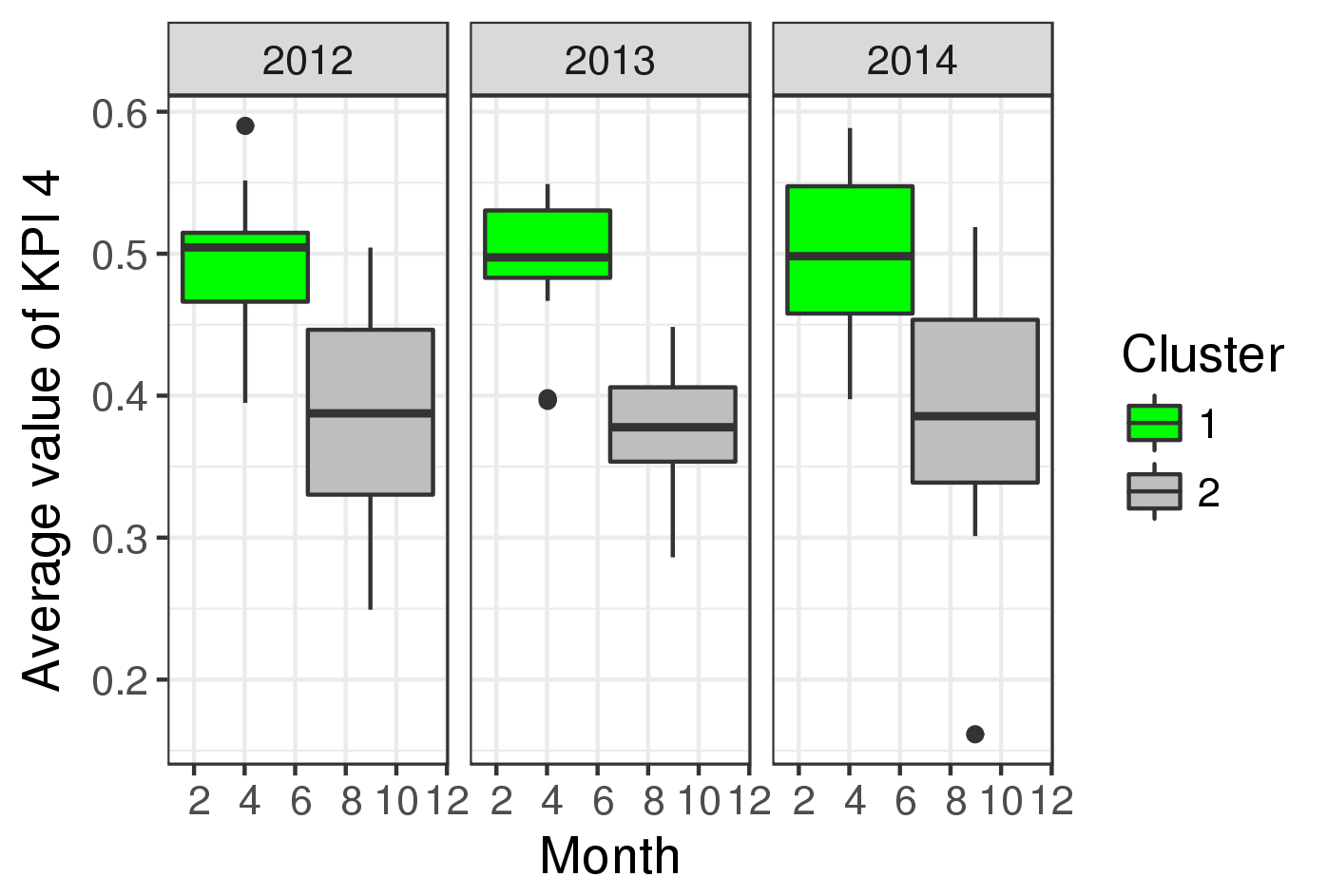}
 \caption{}
 \label{sfig:clust:box2}
\end{subfigure}
\caption[Box Plot for assessing the clusters formed by GMR]{Box Plot for assessing the clusters formed by GMR in the KPI space. Values for each month and year is displayed for Cluster 1 (green) and Cluster 2 (gray) (a) KPI \#3 (b) KPI \#4}
\label{fig:clust:box}
\end{figure}

As shown in Figure~\ref{fig:clust:eval}, the average value of KPI \#1 in Figure~\ref{sfig:clust:eval1} is clearly different between the two clusters and cluster 2 (red) tends to contain the dealers that have a smaller value in that particular KPI. In the case of KPI \#2 in Figure~\ref{sfig:clust:eval2}, there are some months that the two clusters have overlapped, but the two clusters still seem to be different. Figure~\ref{fig:clust:box} shows box plots for two other important KPIs, separated by the clusters. This Figure also proves the effectiveness of GMR in forming clusters with members with different KPI ranges.

\section{Conclusion}\label{sec:conc}
In this study, we introduced a solution to the mixture of regressions problem with observation group structure, which we term Grouped Mixture of Regressions (GMR). We derived the EM updates for this model providing a very fast algorithm for fitting the model. In addition, by deriving the predictive density we showed how the knowledge of the group membership of the new observations improves the prediction in the GMR versus the usual mixture of regressions. 

Monte Carlo simulation experiments confirm the robustness of the algorithm and improved predictive performance. Using cross-validation, GMR successfully selected the optimal number of components which, in general, is a challenging task for any clustering technique. In addition, we performed an empirical study to compare the GMR with another recent heuristic algorithm proposed by \cite{haidar:inpress}, namely \mmclpp. The experiments suggest that GMR outperforms \mmclpp in both the recovery of the true clusters as well as the prediction accuracy. We also demonstrated the effectiveness of the algorithm in a real-world problem (predicting automotive dealership performance) and confirmed the superior performance of the GMR relative to the \mmclpp in this real-world setting.

There are several avenues for potential future research. The current version of the GMR assumes that the covariates (features) are deterministic. This assumption can be relaxed by considering the model that treats the covariates as random. It is also possible to extend the approach to Generalized Linear Models (GLM) setting to expand the potential range of applications.

\small
\bibliographystyle{chicago}
\bibliography{bibfile}

\appendix 

\section{Appendix: EM Updates for Algorithm~\ref{alg:gmr}}\label{sec:EM:details}

Expanding the expected log-likelihood~\eqref{eq:exp_lik} using the definition of~$\gamma_{rk}(\theta)$ in~\eqref{eq:likelihod}, we have
\begin{align}\label{eq:expanded:expec:likelihood}
	F(\theta;\thh) = E_{z \sim \tau(\thh)} [\ell(\theta;z)]
	&=\sum_{k=1}^K \tau_{+k}(\thh) \log \pi_k + 
	\sum_{r=1}^R \sum_{k=1}^K \sum_{i=1}^{n_r}  \tau_{rk}(\thh) \log \phi_{\sigma_k }\big( y_{ri} - \beta_k^T x_{ri} \big).
\end{align}
where $\phi_\sigma(t) := (2\pi \sigma^2)^{-1/2} \exp(-\frac12 t^2/\sigma^2)$ is the density of $N(0,\sigma^2)$.

We would like to maximize~\eqref{eq:expanded:expec:likelihood} over $\theta$. Recall that $\beta_k,x_{ri} \in \reals^p$ where $p$ is the number of features.
We will use $\doteq_\pi$ for example, when the two sides are equal up to additive constants, as functions of $\pi$. Fixing everything and maximizing over $\pi = (\pi_1,\dots,\pi_k)$, we are maximizing $\pi \mapsto \sum_k \tau_{+k}(\thh) \log \pi_k$ over probability vector $\pi$. This is the MLE in the multinomial family and the solution is $\pi_k \propto_k \tau_{+k}$, that is
	\begin{align}\label{eq:pi:update:1}
	    \pi_k = \frac{\tau_{+k}}{\sum_{k'} \tau_{+k'}}  = \frac{\tau_{+k}}{R}
	\end{align}
	where we used $\sum_{k'} \tau_{+k'} = \sum_{k'} \sum_r \tau_{rk'} = \sum_r \sum_{k'} \tau_{rk'} = \sum_r 1 = R$, since for fixed $r$, $\tau_{rk}$ sums to 1 over $k$.
	
	To maximize over $\beta$, we again fix everything else. Since $\log \phi_\sigma(t) \doteq_t -\frac12 (\log \sigma^2 + t^2/\sigma^2) $, we are maximizing
	\begin{align}
	    F(\theta;\thh)\; &\doteq_\beta\; -\sum_r \sum_k \sum_i^{n_r} 
	    \tau_{rk}(\thh) \frac1{2\sigma_k^2} (y_{ri} - \beta_k^T x_{ri})^2 \notag\\
	    &\doteq_\beta\; -\sum_r \sum_k \sum_i^{n_r} 
	    \tau_{rk}(\thh) \frac1{2\sigma_k^2} [(\beta_k^T x_{ri})^2 - 2y_{ri}\beta_k^T x_{ri}] \label{eq:temp:1}
	\end{align}
	ignoring the constant terms generated by $y_{ri}^2$. 
	
	Note that $(\beta_k^T x_{ri})^2 = (\beta_k^T x_{ri})(x_{ri}^T \beta_k) = \beta_k^T (x_{ri}x_{ri}^T) \beta_k$. Similarly, $y_{ri}\beta_k^T x_{ri} = \beta_k^T(y_{ri} x_{ri})$. Let us define
	\begin{align}\label{eq:hat:def}
	    \Sigh_r := \frac1{n_r}\sum_{i=1}^{n_r} x_{ri} x_{ri}^T, \quad 
	    \rhoh_r :=  \frac1{n_r}\sum_{i=1}^{n_r} y_{ri} x_{ri}
	\end{align}
	Summing over $i$ first in~\eqref{eq:temp:1}, we get
	\begin{align}
	     F(\theta;\thh)\; 
	    &\doteq_\beta\; -\sum_r \sum_k
	     \frac{\tau_{rk}}{2\sigma_k^2} n_r[ \beta_k^T \Sigh_r \beta_k - 2 \beta_k^T \rhoh_r] \notag \\
	    &=  - \sum_k \frac1{2\sigma_k^2}
	     \sum_r\tau_{rk} n_r[ \beta_k^T \Sigh_r \beta_k - 2 \beta_k^T \rhoh_r]
	     \label{eq:temp:2}
	\end{align}
	Let us define $w_{rk} :=  n_r \tau_{rk}$ and $\wc_{rk} := w_{rk}/w_{+k}$ where $w_{+k} = \sum_r n_r\tau_{rk}$, and let
	\begin{align}\label{eq:tilde:def}
	    \Sigt_k := \sum_{r=1}^R \wc_{rk} \Sigh_r, \quad 
	    \rhot_k := \sum_{r=1}^R \wc_{rk}  \rhoh_r.
	\end{align}
	Dividing and multiplying by $w_{+k}$ and summing over $r$ in~\eqref{eq:temp:2}, we get
	\begin{align}
	    F(\theta;\thh)
	    \doteq_\beta -\sum_k \frac{w_{+k}}{2\sigma_k^2} [\beta_k^T \Sigt_k \beta_k - 2 \beta_k^T \rhot_k].
	\end{align}
	The problem is separable in $k$, and the minimizer over $\beta_k$ is $\beta_k = \Sigt_k^{-1} \rhot_k$.
	
	\medskip
	To optimize over $\alpha_k:=\sigma_k^2$, let us fix everything else. We have
	\begin{align}\label{eq:temp:495}
	    F(\theta;\thh) \doteq_{\alpha} -\frac12 \sum_k \Big[ \sum_r \sum_i^{n_r} \tau_{rk} \log \alpha_k + \sum_r \sum_i^{n_r} 
	    \tau_{rk}\frac{(y_{ri} - \beta_k^T x_{ri})^2}{\alpha_k}\Big].
	\end{align}
	The first term in brackets is 	 $(\sum_r n_r \tau_{rk}) \log \alpha_k = w_{+k} \log \alpha_k$. Defining
	\begin{align}
	    E_{rk} := E_{rk}(\beta) := \frac1{n_r}\sum_{i}^{n_r}   (y_{ri} - \beta_k^T x_{ri})^2, \qquad 
	    \Eb_k := \Eb_k(\beta) := \sum_r \wc_{rk} E_{rk}.
	\end{align}
	we see that the second term in brackets in~\eqref{eq:temp:495} is just $w_{+k}\Eb_k$. We have
	\begin{align}
	     F(\theta;\thh)   \doteq_{\alpha}
	     -\frac12 \sum_k w_{+k} \Big[  \log \alpha_k + \frac{\Eb_k}{\alpha_k}\Big]
	\end{align}
	This problem is separable in $\alpha_k$ and the solution is $\alpha_k = \Eb_k$. Putting the pieces together, we obtain the Algorithm~\ref{alg:gmr}.

\section{Appendix: Details for $\beta$-error Calculation} \label{sec:avg:beta:derivation}
Let $\Ch_k \subset [R]$ be the $k$th estimated cluster (containing indices of the groups estimated to be in cluster $k$) and $\zh_{r} \in \{0,1\}^K$ the estimated membership vector for group $r$, so that $\zh_{rk} = 1\{r \in \Ch_k\}$. Similarly, let $C_k \subset [R]$ be the true cluster $k$ and $z_{r}$ the true label vector for group $r$, so that $z_{rk} = 1\{z_r \in C_k\}$. The normalized confusion matrix $ \conf = (\conf_{k \ell}) \in [0,1]^{K \times K}$  between the two sets of labels is given by 
$
\conf_{k\ell} = \frac1R \sum_{r=1}^R z_{rk} \; \zh_{r\ell} = \frac1R \sum_{r=1} 1\{r \in C_k, \, r \in \Ch_\ell \}.
$
Then, the following desired result is obtained:
\begin{align*}\label{eq:bet_err}
	\frac1R \sum_{r=1}^R \vnorm{\beh^{(r)} - \beta^{(r)}}^2
	&=	\frac1R \sum_{r=1}^R \Big[
		\sum_{k,\ell=1}^K 1\{r \in C_k, \, r \in \Ch_\ell \}\Big] 
		 \vnorm{\beh^{(r)} - \beta^{(r)}}^2\\
	&=	 \sum_{k,\ell=1}^K \frac1R \sum_{r=1}^R 1\{r \in C_k, \, r \in \Ch_\ell \}
		\vnorm{\beh^{(r)} - \beta^{(r)}}^2\\
	&= \sum_{k,\ell=1}^K \frac1R \sum_{r=1}^R 1\{r \in C_k, \, r \in \Ch_\ell \}
	\vnorm{\beh_\ell - \beta_k}^2\\
	&= \sum_{k,\ell=1}^K \vnorm{\beh_\ell - \beta_k}^2 \frac1R \sum_{r=1}^R 1\{r \in C_k, \, r \in \Ch_\ell \}\\
	&= \sum_{k,r} D_{kr} \conf_{kr} = \tr(D^T \conf)
\end{align*}

\section{Appendix: Experiment \& Results Details}\label{sec:tables}
\begin{figure}[ht] 
	\centering
	\begin{subfigure}[b]{0.32\textwidth}
		\centering
		\includegraphics[width=\textwidth]{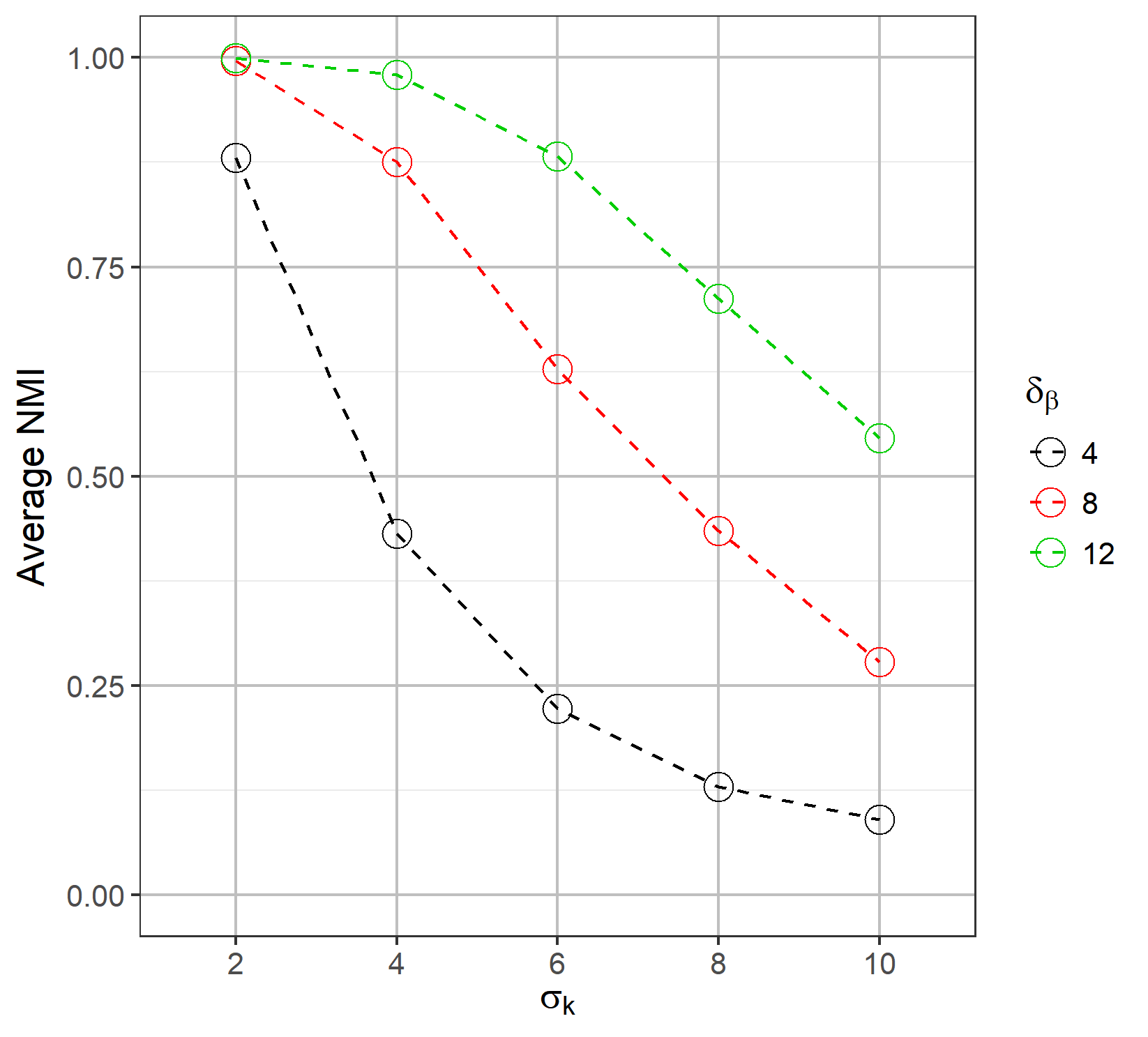}
		\caption{\footnotesize{}}\label{sfig:nmi_n1}
	\end{subfigure}
	\begin{subfigure}[b]{0.32\textwidth}
		\centering
		\includegraphics[width=\textwidth]{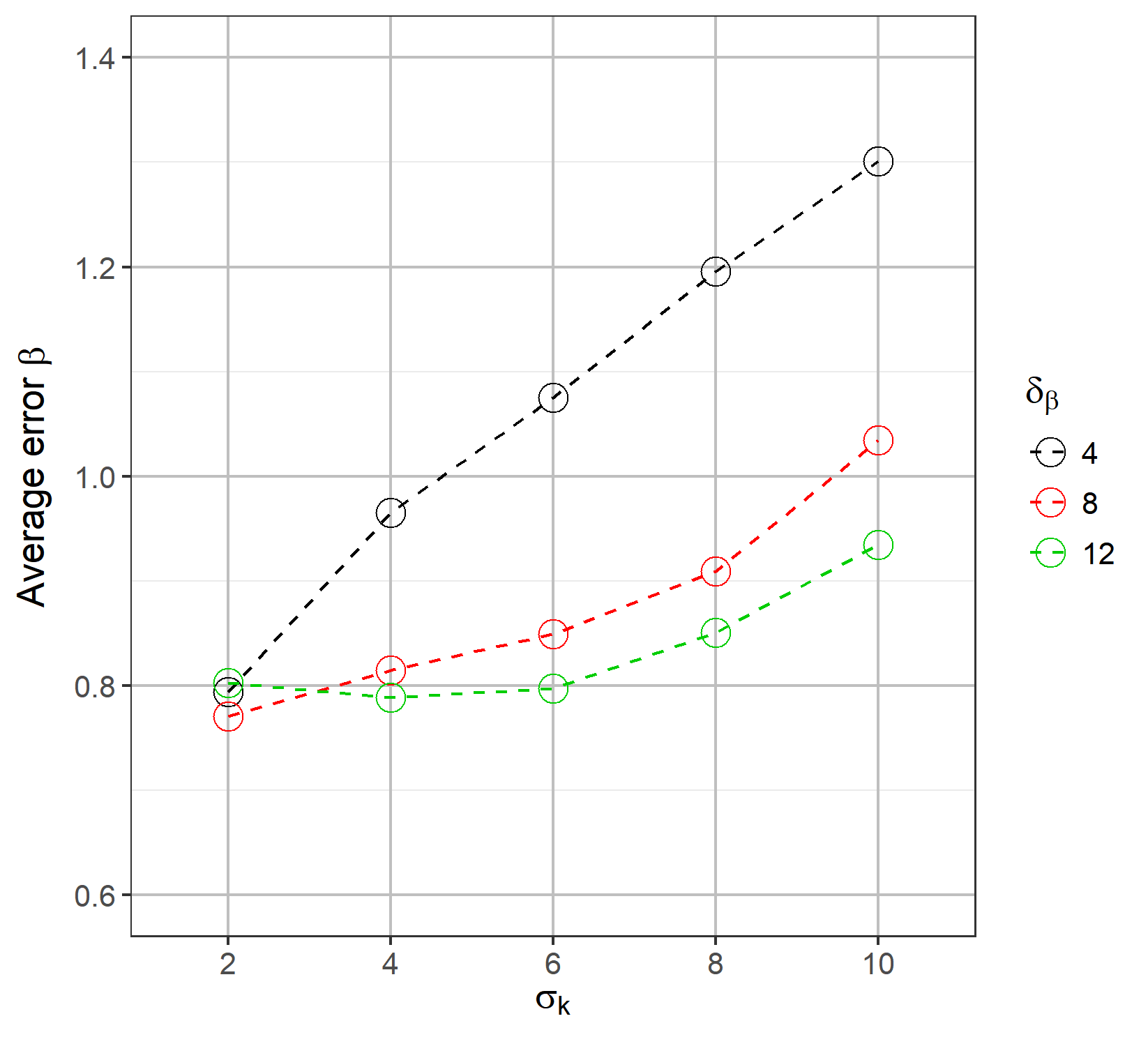}
		\caption{\footnotesize{}}\label{sfig:bt_n1}
	\end{subfigure}
	\begin{subfigure}[b]{0.32\textwidth}
		\centering
		\includegraphics[width=\textwidth]{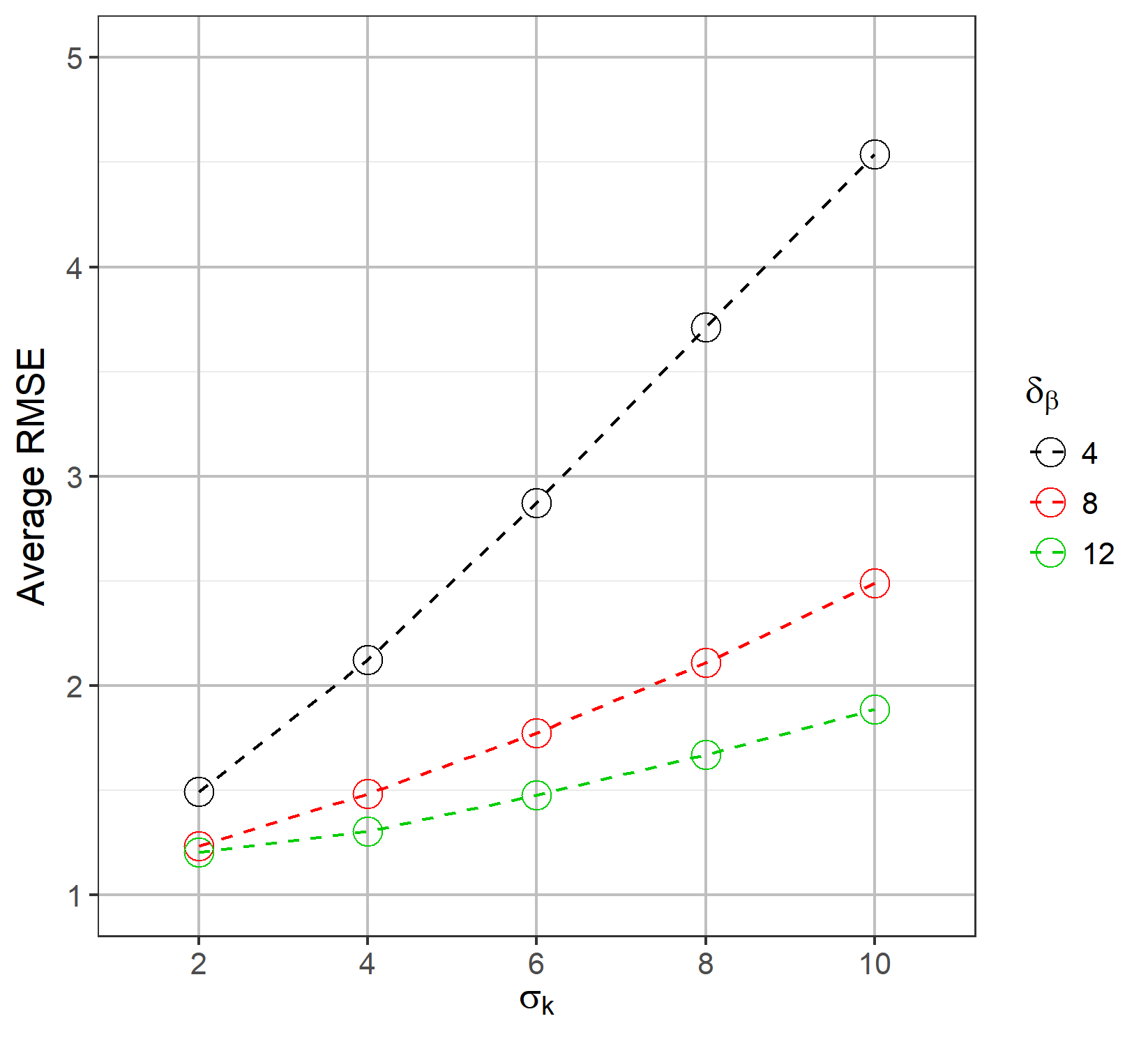}
		\caption{\footnotesize{}}\label{sfig:rmse_n1}
	\end{subfigure}
	\caption[The effect of $\delta_\beta$ and $\sigma_k$ for the case $n=100, K=2, p=2$]{The effect of $\delta_\beta$ and $\sigma_k$ for the case $n=100, K=2, p=2$; each colored line in a plot represents different value of $\delta_\beta$, $X$ axis shows different values of $\sigma_k$, and $y$ axis shows: (a) average NMI, (b) average $\beta$ estimation error, (c) average RMSE for prediction}\label{fig:bet:noise}
\end{figure}

\begin{figure} [ht]
	\centering
	\begin{subfigure}[b]{0.32\textwidth}
		\centering
		\includegraphics[width=\textwidth]{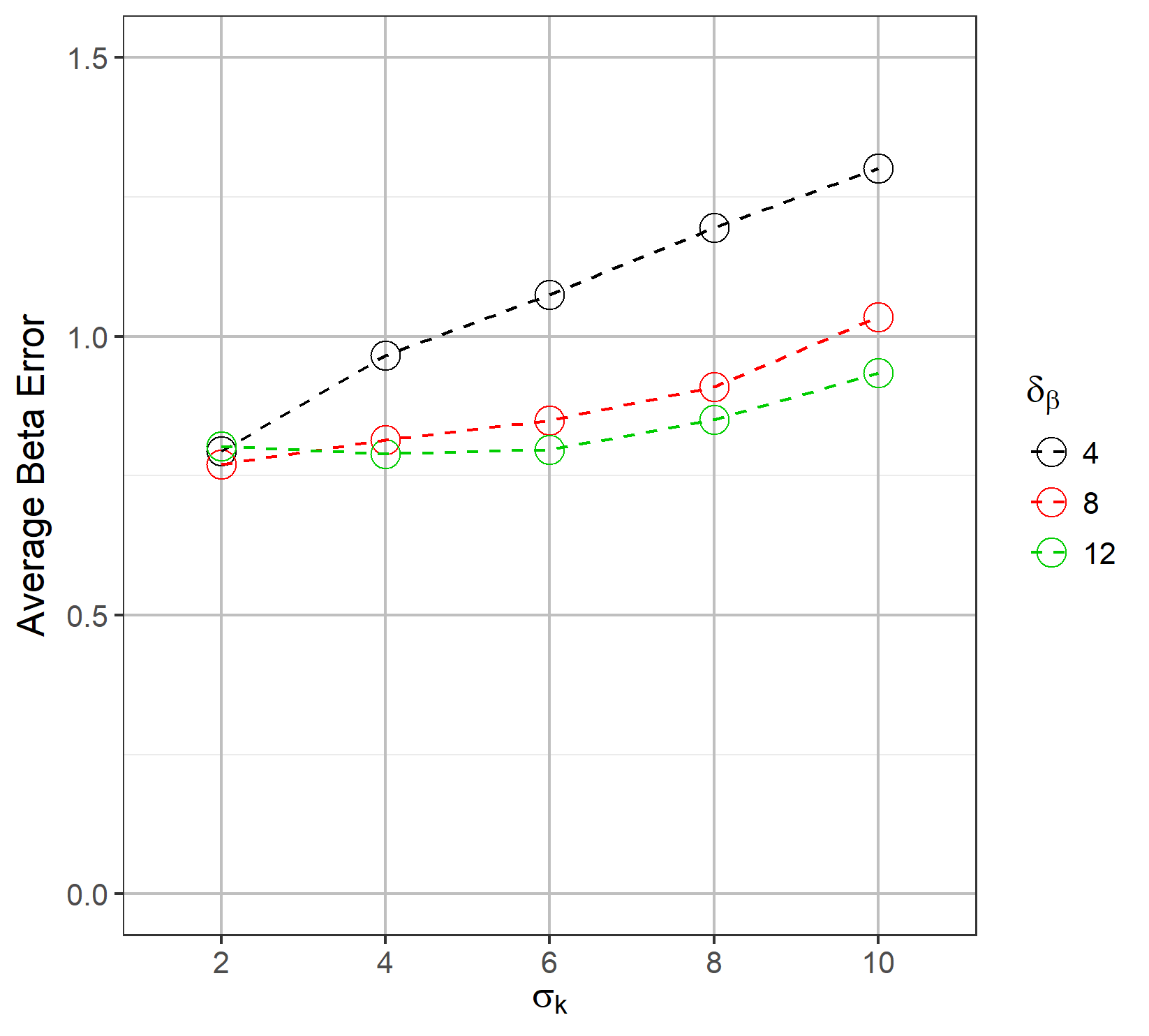}
		\caption{\footnotesize{}}\label{sfig:bet_pk1}
	\end{subfigure}
	\begin{subfigure}[b]{0.32\textwidth}
		\centering
		\includegraphics[width=\textwidth]{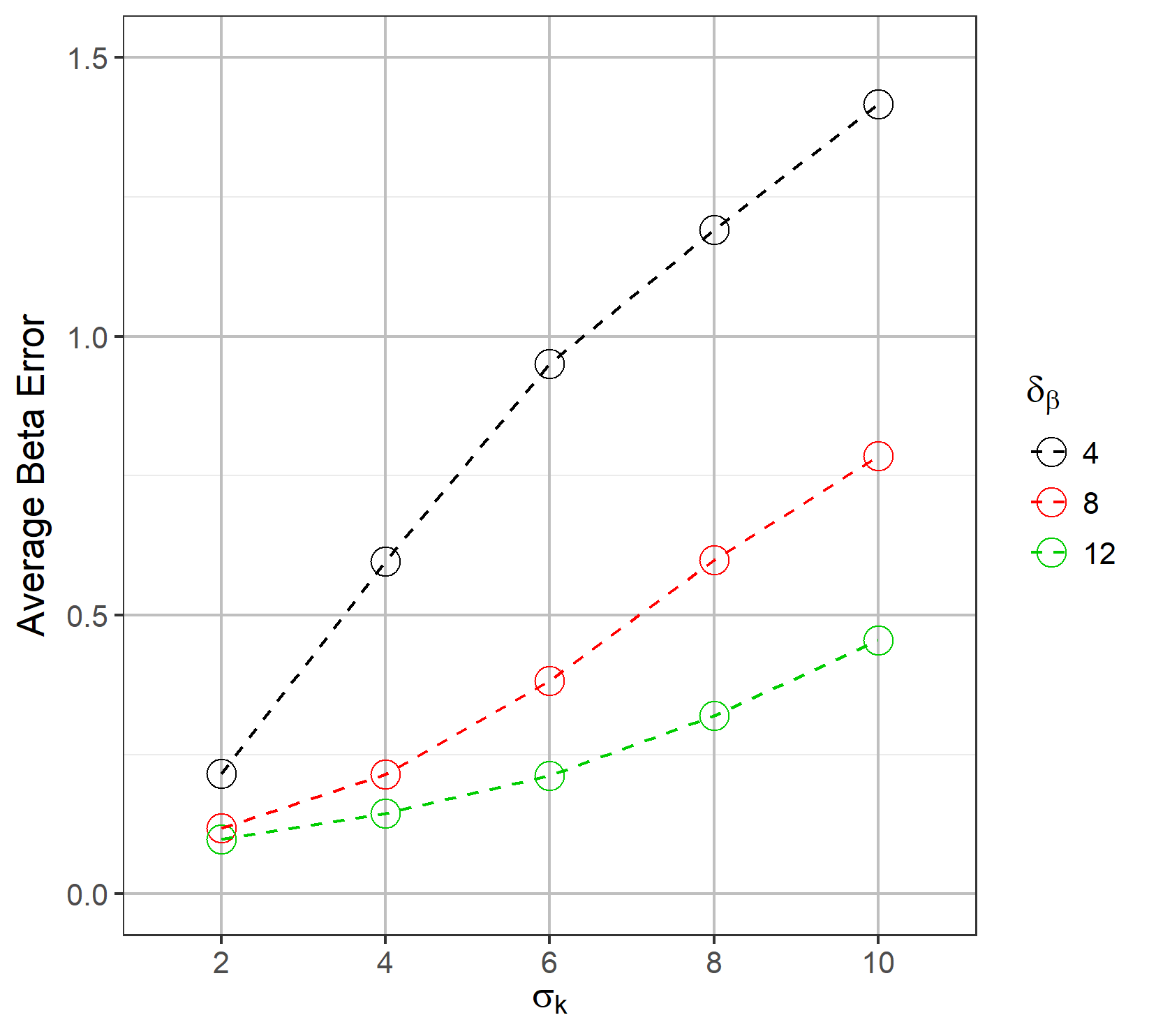}
		\caption{\footnotesize{}}\label{sfig:bet_pk2}
	\end{subfigure}
	\begin{subfigure}[b]{0.32\textwidth}
		\centering
		\includegraphics[width=\textwidth]{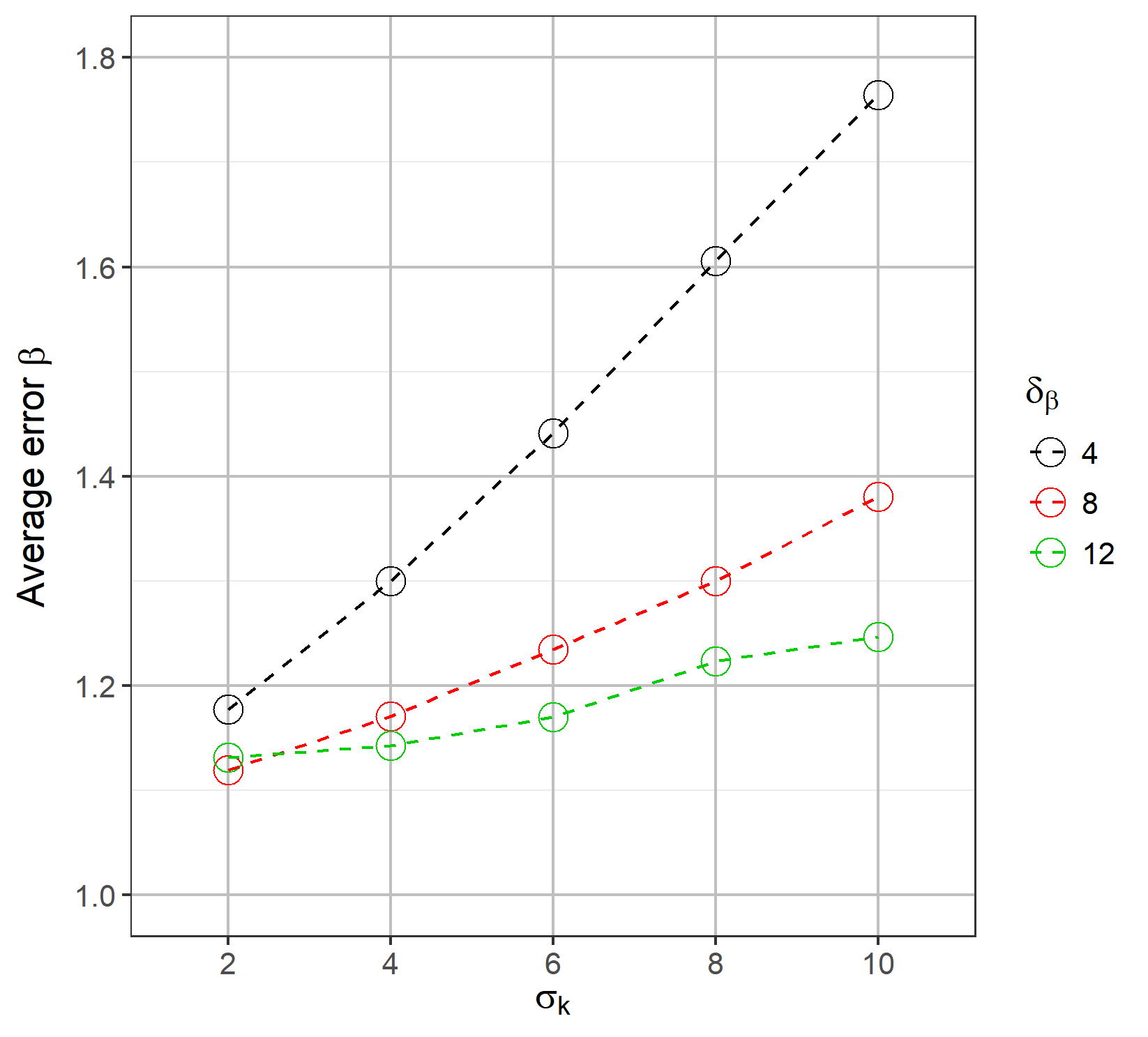}
		\caption{\footnotesize{}}\label{sfig:bet_pk3}
	\end{subfigure}
	\caption[The impact of $K$ and $p$ on $\beta$ estimation error for the case $n=100$]{The impact of $K$ and $p$ on $\beta$ estimation error for the case $n=100$; each colored line in a plot represents different value of $\delta_\beta$, $X$ axis shows different values of $\sigma_k$, and $y$ axis is average NMI for: (a) $K=2$, $p=2$, (b) $K=2$, $p=4$, (c) $K=4$, $p=4$}\label{fig:p:k:bet}
\end{figure}

Figure \ref{fig:p:k:bet} illustrates the impact of $K$ and $p$ on $\beta$ estimation error. By comparing the plots in Figure \ref{fig:p:k:bet}, it is hard to find a consistent pattern for the behavior of the $\beta$ estimation error with respect to $p$ and $K$. What could be noticed is that in the case where $K=2$ and $p=2$, the error is less sensitive to increasing the noise ($\sigma_k$). However, the error stays higher when the noise is smaller (between 2-6). In the case of $\sigma_k=10$, the highest error belongs to the case $K=4$, $p=4$.

\begin{table}[]
\centering
\caption{NMI Performance}
\label{tab:nmi}
\footnotesize
\setlength\tabcolsep{1.5mm}
\begin{tabular}{@{}cc|ccccc|ccccc|ccccc|@{}}
\cmidrule(l){3-17}
                                                     &                  & \multicolumn{5}{c|}{$\delta_\beta$ = 4}                          & \multicolumn{5}{c|}{$\delta_\beta$ = 7}                          & \multicolumn{5}{c|}{$\delta_\beta$ = 11}                         \\ \cmidrule(l){2-17} 
\multicolumn{1}{c|}{}                                & {\textbf{\backslashbox
                                                    {\textit{N}} {$\sigma_k$}}} & \textbf{2} & \textbf{4} & \textbf{6} & \textbf{8} & \textbf{10} & \textbf{2} & \textbf{4} & \textbf{6} & \textbf{8} & \textbf{10} & \textbf{2} & \textbf{4} & \textbf{6} & \textbf{8} & \textbf{10} \\ \midrule
\multicolumn{1}{|c|}{\multirow{4}{*}{\rotatebox{90}{$K$=2; $d$=2}}} & \textbf{100}     & 0.88       & 0.43       & 0.22       & 0.13       & 0.09        & 0.99       & 0.87       & 0.62       & 0.43       & 0.27        & 0.99       & 0.98       & 0.88       & 0.71       & 0.54        \\
\multicolumn{1}{|c|}{}                               & \textbf{200}     & 0.98       & 0.71       & 0.40       & 0.25       & 0.15        & 0.99       & 0.98       & 0.88       & 0.68       & 0.54        & 1          & 0.99       & 0.97       & 0.92       & 0.82        \\
\multicolumn{1}{|c|}{}                               & \textbf{400}     & 0.99       & 0.91       & 0.68       & 0.46       & 0.32        & 1          & 0.99       & 0.98       & 0.90       & 0.81        & 1          & 1          & 0.99       & 0.98       & 0.96        \\
\multicolumn{1}{|c|}{}                               & \textbf{800}     & 1          & 0.99       & 0.98       & 0.90       & 0.75        & 1          & 1          & 1          & 0.99       & 0.99        & 1          & 1          & 1          & 1          & 1           \\ \midrule
\multicolumn{1}{|c|}{\multirow{4}{*}{\rotatebox{90}{$K$=2; $d$=4}}} & \textbf{100}     & 0.93       & 0.49       & 0.21       & 0.13       & 0.09        & 0.99       & 0.93       & 0.73       & 0.48       & 0.32        & 0.99       & 0.99       & 0.93       & 0.8        & 0.64        \\
\multicolumn{1}{|c|}{}                               & \textbf{200}     & 0.99       & 0.82       & 0.45       & 0.26       & 0.16        & 1          & 0.99       & 0.94       & 0.80       & 0.62        & 1          & 0.99       & 0.99       & 0.97       & 0.91        \\
\multicolumn{1}{|c|}{}                               & \textbf{400}     & 1          & 0.97       & 0.79       & 0.54       & 0.35        & 1          & 1          & 0.99       & 0.97       & 0.89        & 1          & 1          & 1          & 0.99       & 0.99        \\
\multicolumn{1}{|c|}{}                               & \textbf{800}     & 1          & 0.99       & 0.96       & 0.84       & 0.64        & 1          & 1          & 1          & 0.99       & 0.98        & 1          & 1          & 1          & 1          & 0.99        \\ \midrule
\multicolumn{1}{|c|}{\multirow{4}{*}{\rotatebox{90}{$K$=4; $d$=4}}} & \textbf{100}     & 0.80       & 0.34       & 0.20       & 0.15       & 0.12        & 0.97       & 0.80       & 0.52       & 0.34       & 0.25        & 0.98       & 0.94       & 0.81       & 0.62       & 0.45        \\
\multicolumn{1}{|c|}{}                               & \textbf{200}     & 0.95       & 0.61       & 0.33       & 0.21       & 0.17        & 0.97       & 0.95       & 0.81       & 0.61       & 0.44        & 0.97       & 0.97       & 0.95       & 0.86       & 0.75        \\
\multicolumn{1}{|c|}{}                               & \textbf{400}     & 0.96       & 0.86       & 0.58       & 0.38       & 0.27        & 0.96       & 0.96       & 0.94       & 0.86       & 0.72        & 0.96       & 0.96       & 0.96       & 0.95       & 0.92        \\
\multicolumn{1}{|c|}{}                               & \textbf{800}     & 0.95       & 0.95       & 0.84       & 0.64       & 0.47        & 0.95       & 0.95       & 0.96       & 0.95       & 0.92        & 0.96       & 0.95       & 0.96       & 0.96       & 0.96        \\ \bottomrule
\end{tabular}
\end{table}

\begin{table}[]
\centering
\caption{$\beta$ Error}
\label{tab:beta}
\footnotesize
\setlength\tabcolsep{1.5mm}
\begin{tabular}{@{}ll|lllll|lllll|lllll|@{}}
\cmidrule(l){3-17}
                                                     &                                       & \multicolumn{5}{c|}{$\beta$-distance = 4}                          & \multicolumn{5}{c|}{$\beta$-distance = 7}                          & \multicolumn{5}{c|}{$\beta$-distance = 11}                         \\ \cmidrule(l){2-17} 
\multicolumn{1}{l|}{}                                & {\textbf{\backslashbox
                                                    {\textit{N}} {$\sigma_k$}}} & \textbf{2} & \textbf{4} & \textbf{6} & \textbf{8} & \textbf{10} & \textbf{2} & \textbf{4} & \textbf{6} & \textbf{8} & \textbf{10} & \textbf{2} & \textbf{4} & \textbf{6} & \textbf{8} & \textbf{10} \\ \midrule
\multicolumn{1}{|l|}{\multirow{4}{*}{\rotatebox{90}{$K$=2; $d$=2}}} & \textbf{100}                          & 0.79       & 0.96       & 1.07       & 1.19       & 1.42        & 0.77       & 0.81       & 0.85       & 0.90       & 1.03        & 0.8        & 0.78       & 0.79       & 0.85       & 0.93        \\
\multicolumn{1}{|l|}{}                               & \textbf{200}                          & 0.76       & 0.81       & 0.97       & 1.06       & 1.15        & 0.78       & 0.81       & 0.79       & 0.87       & 0.90        & 0.79       & 0.76       & 0.78       & 0.81       & 0.81        \\
\multicolumn{1}{|l|}{}                               & \textbf{400}                          & 0.79       & 0.76       & 0.85       & 0.96       & 1.02        & 0.79       & 0.79       & 0.8        & 0.82       & 0.85        & 0.78       & 0.79       & 0.76       & 0.81       & 0.79        \\
\multicolumn{1}{|l|}{}                               & \textbf{800}                          & 0.07       & 0.1        & 0.14       & 0.21       & 0.31        & 0.06       & 0.07       & 0.08       & 0.1        & 0.12        & 0.06       & 0.06       & 0.07       & 0.08       & 0.09        \\ \midrule
\multicolumn{1}{|l|}{\multirow{4}{*}{\rotatebox{90}{$K$=2; $d$=4}}} & \textbf{100}                          & 0.21       & 0.59       & 0.95       & 1.19       & 1.41        & 0.11       & 0.59       & 0.38       & 0.59       & 0.78        & 0.09       & 0.14       & 0.21       & 0.32       & 0.45        \\
\multicolumn{1}{|l|}{}                               & \textbf{200}                          & 0.14       & 0.32       & 0.63       & 0.88       & 1.06        & 0.09       & 0.15       & 0.21       & 0.32       & 0.47        & 0.08       & 0.11       & 0.14       & 0.18       & 0.25        \\
\multicolumn{1}{|l|}{}                               & \textbf{400}                          & 0.11       & 0.19       & 0.35       & 0.55       & 0.76        & 0.08       & 0.10       & 0.14       & 0.19       & 0.26        & 0.07       & 0.09       & 0.11       & 0.13       & 0.16        \\
\multicolumn{1}{|l|}{}                               & \textbf{800}                          & 0.09       & 1.40       & 0.20       & 0.31       & 0.46        & 0.07       & 0.09       & 0.11       & 0.13       & 0.17        & 0.06       & 0.07       & 0.09       & 0.10       & 0.12        \\ \midrule
\multicolumn{1}{|l|}{\multirow{4}{*}{\rotatebox{90}{$K$=4; $d$=4}}} & \textbf{100}                          & 1.17       & 1.3        & 1.44       & 1.60       & 1.76        & 1.12       & 1.17       & 1.23       & 1.3        & 1.38        & 1.13       & 1.14       & 1.41       & 1.62       & 1.83        \\
\multicolumn{1}{|l|}{}                               & \textbf{200}                          & 1.14       & 1.23       & 1.31       & 1.40       & 1.51        & 1.14       & 1.15       & 1.17       & 1.20       & 1.26        & 1.13       & 1.12       & 1.15       & 1.16       & 1.19        \\
\multicolumn{1}{|l|}{}                               & \textbf{400}                          & 1.14       & 1.13       & 1.21       & 1.29       & 1.36        & 1.13       & 1.13       & 1.16       & 1.16       & 1.19        & 1.13       & 1.14       & 1.13       & 1.14       & 1.15        \\
\multicolumn{1}{|l|}{}                               & \textbf{800}                          & 1.13       & 1.14       & 1.16       & 1.21       & 1.25        & 1.13       & 1.13       & 1.14       & 1.14       & 1.14        & 1.14       & 1.13       & 1.13       & 1.14       & 1.14        \\ \bottomrule
\end{tabular}
\end{table}

\begin{table}[]
\centering
\caption{RMSE Performance}
\label{tab:rmse}
\footnotesize
\setlength\tabcolsep{1.5mm}
\begin{tabular}{@{}ll|lllll|lllll|lllll|@{}}
\cmidrule(l){3-17}
                                                     &                                       & \multicolumn{5}{c|}{$\beta$-distance = 4}                       & \multicolumn{5}{c|}{$\beta$-distance = 7}                       & \multicolumn{5}{c|}{$\beta$-distance = 11}                      \\ \cmidrule(l){2-17} 
\multicolumn{1}{l|}{}                                & {\textbf{\backslashbox
                                                    {\textit{N}} {$\sigma_k$}}} & \textbf{2} & \textbf{4} & \textbf{6} & \textbf{8} & \textbf{10} & \textbf{2} & \textbf{4} & \textbf{6} & \textbf{8} & \textbf{10} & \textbf{2} & \textbf{4} & \textbf{6} & \textbf{8} & \textbf{10} \\ \midrule
\multicolumn{1}{|l|}{\multirow{4}{*}{\rotatebox{90}{$K$=2; $d$=2}}} & \textbf{100}                          & 1.49       & 2.12       & 2.87       & 3.71       & 4.54        & 1.23       & 1.48       & 1.77       & 2.10       & 2.49        & 1.2        & 1.30       & 1.47       & 1.67       & 1.88        \\
\multicolumn{1}{|l|}{}                               & \textbf{200}                          & 1.23       & 2.11       & 2.87       & 3.69       & 4.51        & 1.23       & 1.47       & 1.76       & 2.10       & 2.49        & 1.20
       & 1.32       & 1.46       & 1.66       & 1.88        \\
\multicolumn{1}{|l|}{}                               & \textbf{400}                          & 1.46       & 2.11       & 2.88       & 3.68       & 4.52        & 1.25       & 1.48       & 1.78       & 2.13       & 2.48        & 1.23       & 1.32       & 1.46       & 1.66       & 1.89        \\
\multicolumn{1}{|l|}{}                               & \textbf{800}                          & 1.39       & 2.06       & 2.83       & 3.6        & 4.49        & 1.18       & 1.4        & 1.7        & 2.06       & 2.44        & 1.13       & 1.26       & 1.4        & 1.6        & 1.833       \\ \midrule
\multicolumn{1}{|l|}{\multirow{4}{*}{\rotatebox{90}{$K$=2; $d$=4}}} & \textbf{100}                          & 1.45       & 2.10       & 2.86       & 3.69       & 4.53        & 1.23       & 1.45       & 1.75       & 2.11       & 2.50        & 1.18       & 1.30       & 1.44       & 1.65       & 1.86        \\
\multicolumn{1}{|l|}{}                               & \textbf{200}                          & 1.45       & 2.10       & 2.87       & 3.68       & 4.51        & 1.23       & 1.45       & 1.75       & 2.10       & 2.48        & 1.19       & 1.30       & 1.45       & 1.64       & 1.86        \\
\multicolumn{1}{|l|}{}                               & \textbf{400}                          & 0.14       & 2.10       & 2.86       & 3.67       & 4.51        & 1.24       & 1.45       & 1.75       & 2.10       & 2.47        & 1.19       & 1.30       & 1.45       & 1.64       & 1.86        \\
\multicolumn{1}{|l|}{}                               & \textbf{800}                          & 1.44       & 2.09       & 2.86       & 3.67       & 4.50        & 1.23       & 1.45       & 1.74       & 2.09       & 2.46        & 1.19       & 1.29       & 1.45       & 1.65       & 1.86        \\ \midrule
\multicolumn{1}{|l|}{\multirow{4}{*}{\rotatebox{90}{$K$=4; $d$=4}}} & \textbf{100}                          & 1.41       & 2.07       & 2.85       & 3.67       & 4.50        & 1.18       & 1.41       & 1.72       & 2.07       & 2.45        & 1.14       & 1.25       & 1.41       & 1.62       & 1.83        \\
\multicolumn{1}{|l|}{}                               & \textbf{200}                          & 1.41       & 2.06       & 2.84       & 3.66       & 4.49        & 1.19       & 1.40       & 1.72       & 2.06       & 2.45        & 1.15       & 1.26       & 1.41       & 1.61       & 1.82        \\
\multicolumn{1}{|l|}{}                               & \textbf{400}                          & 1.41       & 2.07       & 2.84       & 3.66       & 4.50        & 1.19       & 1.41       & 1.72       & 2.06       & 2.44        & 1.14       & 1.25       & 1.41       & 1.60       & 1.83        \\
\multicolumn{1}{|l|}{}                               & \textbf{800}                          & 1.41       & 2.06       & 2.84       & 3.65       & 4.49        & 1.19       & 1.41       & 1.72       & 2.07       & 2.44        & 1.14       & 1.25       & 1.41       & 1.60       & 1.82        \\ \bottomrule
\end{tabular}
\end{table}

\begin{table}[]
\centering
\caption{Number of Iterations}
\label{tab:itr}
\footnotesize
\setlength\tabcolsep{1.5mm}
\begin{tabular}{@{}ll|lllll|lllll|lllll|@{}}
\cmidrule(l){3-17}
                                                     &                                       & \multicolumn{5}{c|}{$\beta$-distance = 4}                       & \multicolumn{5}{c|}{$\beta$-distance = 7}                                             & \multicolumn{5}{c|}{$\beta$-distance = 11}                      \\ \cmidrule(l){2-17} 
\multicolumn{1}{l|}{}                                & {\textbf{\backslashbox
                                                    {\textit{N}} {$\sigma_k$}}} & \textbf{2} & \textbf{4} & \textbf{6} & \textbf{8} & \textbf{10} & \textbf{2} & \textbf{4} & \textbf{6} & \textbf{8} & \multicolumn{1}{l|}{\textbf{10}} & \textbf{2} & \textbf{4} & \textbf{6} & \textbf{8} & \textbf{10} \\ \midrule
\multicolumn{1}{|l|}{\multirow{4}{*}{\rotatebox{90}{$K$=2; $d$=2}}} & \textbf{100}                          & 14.1       & 53.7       & 84.2       & 100.2      & 109.0       & 4.9        & 14.3       & 32.9       & 53.7       & \multicolumn{1}{l|}{71.3}        & 3.8        & 7.5        & 13.6       & 27.2       & 40.1        \\
\multicolumn{1}{|l|}{}                               & \textbf{200}                          & 6.3        & 27.6       & 60.6       & 82.4       & 101.3       & 3.3        & 6.4        & 14.4       & 29.3       & \multicolumn{1}{l|}{44.7}        & 3.1        & 3.8        & 6.6        & 11.5       & 19.6        \\
\multicolumn{1}{|l|}{}                               & \textbf{400}                          & 3.6        & 12.3       & 33.2       & 55.4       & 74.1        & 2.9        & 3.6        & 6.4        & 12.9       & \multicolumn{1}{l|}{20.4}        & 2.8        & 3.0        & 3.5        & 5.6        & 8.3         \\
\multicolumn{1}{|l|}{}                               & \textbf{800}                          & 2.7        & 3.4        & 6.8        & 13.9       & 24.7        & 2.5        & 2.8        & 3          & 3.4        & \multicolumn{1}{l|}{4.7}         & 2.4        & 2.6        & 2.8        & 2.9        & 3.1         \\ \midrule
\multicolumn{1}{|l|}{\multirow{4}{*}{\rotatebox{90}{$K$=2; $d$=4}}} & \textbf{100}                          & 11.4       & 42.8       & 65.1       & 72.4       & 73.6        & 4.2        & 12.1       & 25.2       & 43.2       & \multicolumn{1}{l|}{55.2}        & 3.6        & 5.8        & 12.0       & 19.6       & 30.8        \\
\multicolumn{1}{|l|}{}                               & \textbf{200}                          & 4.8        & 20.6       & 45.8       & 65.4       & 73.4        & 3.2        & 4.8        & 11.2       & 20.5       & \multicolumn{1}{l|}{32.8}        & 3.1        & 3.4        & 4.7        & 8.2        & 13.7        \\
\multicolumn{1}{|l|}{}                               & \textbf{400}                          & 3.2        & 8.8        & 23.3       & 42.5       & 57.5        & 2.9        & 3.2        & 4.7        & 8.6        & \multicolumn{1}{l|}{15.7}        & 2.8        & 3.0        & 3.2        & 3.8        & 5.7         \\
\multicolumn{1}{|l|}{}                               & \textbf{800}                          & 2.8        & 4.0        & 9.5        & 19.7       & 34.2        & 2.5        & 2.8        & 3.1        & 3.8        & \multicolumn{1}{l|}{6.1}         & 2.4        & 2.6        & 2.8        & 3.1        & 3.3         \\ \midrule
\multicolumn{1}{|l|}{\multirow{4}{*}{\rotatebox{90}{$K$=4; $d$=4}}} & \textbf{100}                          & 41.1       & 114.0      & 149.7      & 163.2      & 171.1       & 13.0       & 40.5       & 81.1       & 113.4      & \multicolumn{1}{l|}{134.1}       & 8.5        & 20.1       & 39.3       & 67.9       & 94.1        \\
\multicolumn{1}{|l|}{}                               & \textbf{200}                          & 18.6       & 74.7       & 121.8      & 149.1      & 162.5       & 7.7        & 18.3       & 42.7       & 72.8       & \multicolumn{1}{l|}{103.0}       & 9.15       & 10.7       & 18.5       & 33.2       & 52.3        \\
\multicolumn{1}{|l|}{}                               & \textbf{400}                          & 11.6       & 36.5       & 81.5       & 117.1      & 142.7       & 11.3       & 12.3       & 20.2       & 36.2       & \multicolumn{1}{l|}{59.8}        & 11.0       & 11.8       & 12.8       & 17.1       & 25.4        \\
\multicolumn{1}{|l|}{}                               & \textbf{800}                          & 12.7       & 18.1       & 40.5       & 72.8       & 104.6       & 13.0       & 13.2       & 12.0       & 17.8       & \multicolumn{1}{l|}{27.7}        & 10.1       & 12.6       & 12.7     & 12.5      & 14.0        \\ \bottomrule
\end{tabular}
\end{table}


%
%
\end{document}